# Ground/space, passive/active remote sensing observations coupled with particle dispersion modelling to understand the inter-continental transport of wildfire smoke plumes

M. Sicard[1,2], M. J. Granados-Muñoz[1], L. Alados-Arboledas[3,4], R. Barragán[1,2], A.E. Bedoya-Velásquez[3,4], J.A. Benavent-Oltra[3,4], D. Bortoli[5], A. Comerón[1], C. Córdoba-Jabonero[6], M. J. Costa[5], A. del Águila[6], A. J. Fernández[7], J.L. Guerrero-Rascado[3,4], O. Jorba[8], F. Molero[7], C. Muñoz-Porcar[1], P. Ortiz-Amezcua[3,4], N. Papagiannopoulos[1,9], M. Potes[5], M. Pujadas[7], F. Rocadenbosch[1,2], A. Rodríguez-Gómez[1], R. Román[10], R. Salgado[5], V. Salgueiro[5], Y. Sola[11], M. Yela[6]

[1]Remote Sensing Laboratory / CommSensLab, Universitat Politècnica de Catalunya, Barcelona, 08034, Spain

[2]Ciències i Tecnologies de l'Espai - Centre de Recerca de l'Aeronàutica i de l'Espai / Institut d'Estudis Espacials de Catalunya (CTE-CRAE / IEEC), Universitat Politècnica de Catalunya, Barcelona, 08034, Spain

[3]Department of Applied Physics, University of Granada, Granada, 18071, Spain

[4]Andalusian Institute for Earth System Research (IISTA-CEAMA), Granada, 18006, Spain

[5]Institute of Earth Sciences and Dept. of Physics, Universidade de Évora, Évora, 7000-671, Portugal

[6]Instituto Nacional de Técnica Aeroespacial (INTA), Atmospheric Research and Instrumentation Branch, Torrejón de Ardoz (Madrid), 28850, Spain

[7]Dept. of Environment, Research Centre for Energy, Environment and Technology (CIEMAT), Madrid, 28040, Spain

[8]Dept. of Earth Sciences, Barcelona Supercomputing Center (BSC), Barcelona, 08034, Spain

[9]Consiglio Nazionale delle Ricerche, Istituto di Metodologie per l'Analisi Ambientale (CNR-IMAA), Tito Scalo, 85050, Italy

[10]Atmospheric Optics Group (GOA), University of Valladolid, 47002, Spain

[11]Dept. of Astronomy and Meteorology, Universitat de Barcelona, Barcelona, 08028, Spain

*Correspondence to*: Michaël Sicard (msicard@tsc.upc.edu)

**Abstract.** During the 2017 record-breaking burning season in Canada / United States, intense wild fires raged during the first week of September in the Pacific northwestern region (British Columbia, Alberta, Washington, Oregon, Idaho, Montana and northern California) burning mostly temperate coniferous forests. The heavy loads of smoke particles emitted in the atmosphere reached the Iberian Peninsula (IP) a few days later on 7 and 8 September. Satellite imagery allows to identify two main smoke clouds emitted during two different periods that were injected and transported in the atmosphere at several altitude levels. Columnar properties on 7 and 8 September at two Aerosol Robotic Network (AERONET) mid-altitude, background sites in northern and southern Spain are: aerosol optical depth (AOD) at 440 nm up to 0.62, Ångström exponent of 1.6-1.7, large dominance of small particles (fine mode fraction > 0.88), low absorption AOD at 440 nm (<0.008) and large single scattering albedo at 440 nm (>0.98). Profiles from the Cloud-Aerosol Lidar with Orthogonal Polarization (CALIOP) show the presence of smoke particles in the stratosphere during the transport, whereas the smoke is only observed in the troposphere at its arrival over the IP. Portuguese and Spanish ground lidar stations from the European Aerosol Research Lidar Network / Aerosols, Clouds, and Trace gases Research InfraStructure Network (EARLINET/ACTRIS) and the Micro-Pulse Lidar NETwork (MPLNET) reveal smoke plumes with different properties: particle depolarization ratio and color ratio, respectively, of 0.05 and 2.5 in the mid troposphere (5 – 9 km) and of 0.10 and 3.0 in the upper troposphere (10 – 13 km). In the mid troposphere the particle depolarization ratio does not seem time-dependent during the transport whereas the color ratio seems to increase (larger particles sediment first). To analyze the horizontal and vertical transport of the smoke from its origin to the IP, particle dispersion modelling is performed with the Hybrid Single Particle Lagrangian Integrated Trajectory Model (HYSPLIT) parameterized with satellite-derived biomass burning emission estimates from the Global Fire Assimilation System (GFAS) of the Copernicus Atmosphere Monitoring Service (CAMS). Three compounds are simulated: carbon monoxide, black carbon and organic carbon. The results show that the first smoke plume which travels slowly reaches rapidly (~1 day) the upper troposphere and lower stratosphere (UTLS) but also shows evidence of large scale horizontal dispersion, while the second plume, entrained by strong subtropical jets, reaches the upper troposphere much slower (~2.5 days). Observations and dispersion modelling all together suggest that particle depolarization properties are enhanced during their vertical transport from the mid to the upper troposphere.

# 1 Introduction

It is well established that atmospheric biomass burning from either prescribed fires or natural wildfires have effects on air quality, atmospheric circulation and climate (Stocks et al., 2003). Wildfires have recently become a focus of growing interest and attention because of their capabilities to inject smoke particles at high altitude levels. The mechanisms leading to the vertical transport of smoke particles are either direct injection by pyroconvection (Fromm et al., 2000; Fromm and Servranckx, 2003), a combination of pyroconvection and radiatively driven uplift forces (de Laat et al., 2012) or a combination of pyroconvection and gravito-photophoresis (Rohatschek, 1996; Pueschel et al., 2000). The first mechanism, pyroconvection, materializes through the formation of pyrocumulus (pyroCu) and their most extreme form, namely pyrocumulonimbus (pyroCb; Fomm et al., 2005). The characteristic injection height of pyroCu and pyroCb emissions is the upper troposphere (UT) and less frequently the lower stratosphere (LS) (Fromm et al., 2010). The second mechanism, called self-lifting, is based on the absorption of incoming solar radiation by soot and smoke particles which may cause sufficient warming for air masses to provide buoyancy and subsequent lofting of the injected plume (Boers et al., 2010; de Laat et al., 2012). The third mechanism, gravito-photophoresis, is due to "a sunlight-induced force acting on particles which are geometrically asymmetric and which have uneven surface distribution of thermal accommodation coefficients" (Pueschel et al., 2000). It is strongly altitude-dependent because of the weak lifting forces involved and it is most effective above 10 km. Renard et al. (2008) suggested that soot from biomass burning could reach the stratosphere owing to the gravito-photophoresis effect. The last two mechanisms, self-lifting and gravito-photophoresis, can only act on particles which are already settled in the free troposphere or in the stratosphere, and thus require a prior injection of the particles usually produced by pyroconvection.

Once in the UT, the tropopause acts as a dynamic barrier to the upward transport of smoke particles from the troposphere because of the steep gradient in the temperature lapse rate, and in most cases the particles stay in the troposphere. The conditions (burnt matter, fire characteristics, latitude range, local meteorology, synoptic conditions, dynamics, etc.) allowing for the penetration of smoke particles through the tropopause are still not yet entirely clear, and many conclusions of the recent literature on the subject call for more investigation on the topic. The transport of particles in the upper troposphere and lower stratosphere (UTLS) has several effects: (i) in this altitude range, the particles can persist for long durations (Robock, 2000), allowing for gradual spread over hemispheric or global scales; (ii) the long-lived aerosol radiative effects, especially marked for smoke which is a warming agent, may cause differential regional heating patterns that affect regional circulation (Lau et al., 2008; Son et al., 2009); (iii) complex interactions with clouds due to their capability to serve as cloud condensation nuclei, producing in the end a reduction of precipitation (see details in Rosenfeld et al., 2007); (iv) effects on UTLS ozone chemistry (Crutzen and Andreae, 1990; Forster et al., 2001; Real et al., 2008). An increasing number of recent studies report on the observation of the presence of smoke particles in the UTLS: Nédélec et al. (2005), Damoah et al. (2006), Rosenfeld et al. (2007), Fromm et al. (2010) (and references therein), Siddaway and Petelina (2011), de Laat et al. (2012), Khaykin et al. (2018), Ansmann et al. (2018), Haarig et al. (2018), and Hu et al. (2018), among others. The number of modelling studies dealing with the injection of smoke into the UTLS is more reduced: Trentmann et al. (2006), Luderer et al. (2006), Cunningham and Reeder (2009), Cammas et al. (2009), and Peterson et al. (2017).

During summer 2017, North America lived one of its worst burning season on record. On 16 August an aerosol index (AI), a qualitative index indicating the presence of elevated layers of aerosols with significant absorption, of 55.4 was recorded over Canada by the Ozone Mapping and Profiling Suite (OMPS) on board Suomi National Polar-orbiting Partnership satellite (Seftor, 2017a). It breaks the record of AI values by far, the previous record being 31.2 registered by the Total Ozone Mapping Spectrometer (TOMS) on 29 May 2001 during the Canadian Chisholm fires (Fromm et al., 2008). The cluster of the most intense fires of August 2017 was located in Canada near the intersection border of Saskatchewan, Alberta and the Northern Territories at latitude 60 °N. These intense fires produced strong pyroCb which injected smoke particles in the LS which travelled eastward, entrained and dispersed zonally by polar jet streams (Khaykin et al., 2018). Smoke layers at 14 – 16 km with an aerosol optical depth (AOD) at 532 nm of 0.6 were observed in Germany (Ansmann et al., 2018) on 22 August. The event is already documented by a series of papers: Khaykin et al. (2018), Ansmann et al. (2018), Haarig et al. (2018), Hu et al. (2018) and Baars et al. (2019). Fifteen days later, on 30 August, AI from OMPS peaked again at 23 in a smoke plume detected over the southern parts of Alberta and Saskatchewan and the upper Great Plains of the United States (US) (Seftor, 2017b). Most of the fires of this new burning period were in the Pacific northwestern region (British Columbia, Alberta, Washington, Oregon, Idaho, Montana and northern California). In the US severe air quality issues were reported in Washington and Oregon at least until 6 September (NYT, 2018). Prevailing winds and the presence of a frontal boundary across the North American continent created the conditions for the formation of a long, wide, arching ribbon of smoke that stretched thousands of kilometers from the source region all the way to Newfoundland, location from where it was further transported towards Europe. The smoke hit the Iberian Peninsula (IP) in southwestern Europe on 7 and 8 September (Sicard et al., 2018). Although the smoke plume was detected in the LS at some points during its transport, it was only detected in the UT over the IP.

This paper investigates the time-space evolution of the smoke plume detected at its arrival over the IP on 7 and 8 September with ground-based multi-wavelength lidars and backward in time with the CALIOP (Cloud-Aerosol Lidar with Orthogonal Polarization) spaceborne lidar, in terms of optical properties and vertical distribution. Sun-sky photometers at mid-altitude, background sites with no local sources are used to monitor the smoke columnar properties over the IP. A dispersion model parameterized with satellite-derived fire products simulates the vertical and horizontal transport of 3 smoke-related compounds: carbon monoxide, black carbon and organic carbon. Simulations, and especially the injection heights computed by the dispersion model, are qualitatively evaluated against observations and used to understand the atmospheric causal mechanisms yielding to the differences observed in the optical properties over the IP in the mid and upper troposphere.

## 2 Instrumentation and tools

The tools used in our methodology include passive/active, ground-based and spaceborne observations, as well as a particle dispersion model. The observations, listed in For each fire simulated, a series of common parameters (in brackets we indicate HYSPLIT denomination) are necessary: location (Release location), start time (Release start time), duration (Release duration) and heat release (Heat release for plume rise). And for each chemical compound simulated (gas or particle), the emission rate (Emission rate) is necessary. Such information is extracted from the biomass burning emission estimates from GFAS (Global Fire Assimilation System; Kaiser et al., 2012) data from CAMS (Copernicus Atmosphere Monitoring Service). GFAS **Table 1**, are used to follow the transport of the smoke plumes from the source to the IP. In addition, passive spaceborne observations are also used to parameterize the emission of the particle dispersion model.

### 2.1 Ground observations

The ground-based observations include lidars and sun-sky photometers in the IP. A total of five lidar systems are used: three from the EARLINET/ACTRIS (European Aerosol Research Lidar Network / Aerosols, Clouds, and Trace Gases Research Infrastructure Network; https://www.actris.eu/default.aspx; Pappalardo et al., 2014) network in Évora (EV), Granada (GR) and Madrid (MA), and two from MPLNET (Micro-Pulse Lidar Network; https://mplnet.gsfc.nasa.gov/; Welton et al., 2001) in El Arenosillo/Huelva (AR) and Barcelona (BA; see For each fire simulated, a series of common parameters (in brackets we indicate HYSPLIT denomination) are necessary: location (Release location), start time (Release start time), duration (Release duration) and heat release (Heat release for plume rise). And for each chemical compound simulated (gas or particle), the emission rate (Emission rate) is necessary. Such information is extracted from the biomass burning emission estimates from GFAS (Global Fire Assimilation System; Kaiser et al., 2012) data from CAMS (Copernicus Atmosphere Monitoring Service). GFAS **Table 1** for more details and Figure 1 for the geographical position of the stations). The EARLINET lidars are multi-wavelength systems measuring at least at three elastic wavelengths. In addition, EV and GR have Raman and depolarization-sensitive channels. The MPLNET systems have one wavelength at 532 nm and an additional polarization-sensitive channel. A review of the lidar techniques using elastic, Raman and depolarization-sensitive channels, among others, for the remote sensing of aerosols can be found in Comerón et al. (2017). For the characterization of the smoke plume, we use the particle depolarization ratio, $\delta_p$, in EV, AR, GR and BA and the pair (color ratio, depolarization ratio) in EV and GR. The particle depolarization ratio and the color ratio provide significant information on the particle shape and dominant size (Burton et al., 2012), respectively. The particle depolarization ratio is defined as (Freudenthaler et al., 2009):

$$\delta_p = \frac{\beta^{\parallel}}{\beta^{\perp}} \qquad (1)$$

where $\beta^{\parallel}$ and $\beta^{\perp}$ are the particle parallel and perpendicular backscatter coefficients, respectively. The color ratio, $CR$, is defined as a function of the particle backscatter coefficient at 532 nm, $\beta_{532}$, and at 1064 nm, $\beta_{1064}$, as:

$$CR = \frac{\beta_{532}}{\beta_{1064}} \qquad (2)$$

The reason for using this definition of the color ratio between the wavelengths of 532 and 1064 nm is that it allows direct comparison with the space-borne lidar (see Section 2.2) and it is a common parameter used in aerosol classification (Burton et al., 2012; Groß et al., 2013). To understand the reasons of the differences and similarities found in the upcoming discussion, we also define the extinction-related Ångström exponent (AE) between the wavelengths of 355 and 532 nm:

$$\alpha - AE = -\ln\left(\frac{\alpha_{355}}{\alpha_{532}}\right) \bigg/ \ln\left(\frac{355}{532}\right) \qquad (3)$$

where $\alpha_{355}$ and $\alpha_{532}$ are the extinction coefficient at 355 and 532 nm, respectively. This quantity is calculated only at EV which is the only stations where Raman inversions were successfully performed. Similarly to the color ratio, $\alpha - AE$ provides information on the particle dominant size. The advantage of $\alpha - AE$ is that it can be directly compared to the Ångström exponent retrieved by AERONET and defined in the next paragraph. While the MPLNET and the EV systems work continuously 24/7, GR and MA measurements are discontinuous.

Due to the high vertical extension of the smoke plume (up to 14 km) and the high AOD values at 440 nm reached at peak (0.6), no Raman inversions could be performed satisfactorily in Granada. Raman inversions performed in Évora yielded a lidar ratio (the extinction-to-backscatter ratio), $LR$, at 532 nm, $LR_{532}$, in the mid troposphere smoke plume on the order of 55 steradian (sr) (see Section 5.2). This value of 55 sr is used in the elastic inversions performed for the other systems at both 532 and 1064 nm. To maximize the signal-to-noise ratio and thus minimize the retrieval uncertainties, all ground-based lidar measurements presented in this work are nighttime measurements. For the EARLINET systems, the Raman-inverted extinction coefficient has an accuracy of 10 – 30 %, the backscatter coefficient of 5 – 10 % and the lidar ratio of 20 – 35 % (Ansmann et al., 2002). As far as elastic inversions are concerned, the uncertainty of the backscatter coefficient is 10 – 20 % according to Ansmann et al. (2002) and the one of the extinction coefficient is almost directly proportional to the uncertainty of the lidar ratio assumed. Thus, a 25% uncertainty in the lidar ratio input parameter (assuming variations of 14 sr around 55 sr, see Section 5.2) of the elastic inversion leads to a relative uncertainty of about 25% in the extinction coefficient. The particle depolarization ratio uncertainty can reach up to 50 % in the UTLS (Rodríguez-Gómez et al., 2017). For the MPLNET systems, according to Córdoba-Jabonero et al. (2018) the backscactter coeffcient and the particle depolarization ratio retrieved from MPL data have a relative uncertainty of 5 to 20 % and of 10 to 60 %, respectively.

In order to monitor the event over the IP from columnar optical properties we looked at mid-altitude AERONET (Aerosol Robotic Network; Holben et al., 1998) sites with no local sources so as to maximize the signature of the smoke long-range transport. Such sites are Montsec in northeastern Spain and Cerro Poyos in south Spain (see For each fire simulated, a series of common parameters (in brackets we indicate HYSPLIT denomination) are necessary: location (Release location), start time (Release start time), duration (Release duration) and heat release (Heat release for plume rise). And for each chemical compound simulated (gas or particle), the emission rate (Emission rate) is necessary. Such information is extracted from the biomass burning emission estimates from GFAS (Global Fire Assimilation System; Kaiser et al., 2012) data from CAMS (Copernicus Atmosphere Monitoring Service). GFAS **Table 1** for more details and Figure 1 for the geographical position of the stations). We considered AERONET Version 3 products: AOD and SDA (Spectral Deconvolution Algorithm; O'Neill et al., 2001; 2003) inversions data level 1.5 in Montsec (level 2.0 is not available yet) and 2.0 in Cerro Poyos; and aerosol inversions data level 1.5 at both sites. The AERONET products used in our work are:

- The AOD at 440 nm, $AOD_{440}$, which has an estimated accuracy of ±0.02 (Eck et al., 1999).

- The Ångström exponent calculated between the wavelengths of 440 and 870 nm, $AE_{440-870}$, which has an accuracy of ±0.25 for $AOD_{440} \geq 0.1$ (Toledano et al., 2007).

- The fine mode fraction, $FMF$, which has an uncertainty of ~25% for an AOD at 500 nm greater than 0.3 (O'Neill et al., 2003). $FMF$ represents the ratio of the fine-mode AOD to the total AOD.

- The absorption aerosol optical depth, $AAOD$, which has an accuracy of ±0.01 for wavelengths greater than 440 nm (Sicard et al., 2016). $AAOD$ represents the AOD due to absorption.

- The single scattering albedo, $SSA$, which has an accuracy of ±0.03 for $AOD_{440} \geq 0.5$ for biomass burning (Sicard et al., 2016). $SSA$ represents the fraction of the AOD due to scattering (i.e. $AOD - AAOD$) to the total AOD.

- The asymmetry factor, $g$, which has an accuracy in the range [±0.03, ±0.08] for biomass burning (Sicard et al., 2016). $g$ represents a measure of the preferred scattering direction and varies between -1 (only backward-scattering, i.e., at 180º relative to the incident direction) and +1 (only forward-scattering at 0º).

## 2.2 Spaceborne observations

Several types of satellite sensors are used to fulfill the objectives of the study. The Atmospheric Infrared Sounder (AIRS; Chahine et al., 2006), on board the Aqua satellite, is a hyperspectral instrument with 2378 infrared channels and 4 visible/near-infrared channels. AIRS, together with the Advanced Microwave Sounding Unit (AMSU-A) and the Humidity Sounder for Brazil (HSB), form the AIRS instrument suite which is designed to measure the Earth's atmospheric water vapor and temperature profiles on a global scale. The physical product from AIRS used in our study is the Carbon Monoxide (CO) Total Column science parameter which is a parameter of the AIRS Level 2 standard retrieval product using AIRS only (AIRS2RET_NRT). It indicates the amount of CO in the vertical column of the atmosphere and is measured in parts per billion by volume (ppbv). The spatial resolution of the AIRS2RET_NRT product is 45 km at nadir. The temporal resolution is twice daily (day and night).

To track back the vertical distribution of the smoke plume before its arrival in the IP, we use the spaceborne lidar Cloud-Aerosol Lidar with Orthogonal Polarization (CALIOP; Winker et al., 2007), on board the Cloud-Aerosol Lidar and Infrared Pathfinder Satellite Observation (CALIPSO) satellite. CALIOP is a two-wavelength polarization-sensitive lidar that provides high-resolution vertical profiles of aerosols and clouds. It utilizes three receiver channels: one measuring the 1064 nm backscatter intensity and two channels measuring orthogonally polarized components of the 532 nm backscattered signal. The data used in our study are the CALIOP Aerosol Profile Lidar Level 2 data, version 4.10. Profiles of extinction and backscatter coefficients at 532 and 1064 nm, as well as particle depolarization ratio at 532 nm, are given at a horizontal resolution of 5 km and a vertical resolution of 60 m. The uncertainty in the aerosol extinction coefficient is 40 % (assumed a 30-% lidar ratio uncertainty) and the one in the aerosol backscatter coefficient is 20 – 30 % at 532 nm (Young et al., 2009). The CALIOP Level 2, version 4.10 data products used in this study contain substantial changes over the earlier releases, among which the most significant is the updated lidar ratio assignment (Young et al., 2018). Information on CALIOP aerosol sub-typing algorithm and assigned lidar ratios can be found in Omar et al. (2018) and Kim at al. (2018). CALIOP observations have been used for the study of long-range transport of fire smoke locally (Kar et al., 2018) and also globally (Mehta and Singh, 2018).

The Moderate Resolution Imaging Spectroradiometer (MODIS; Kaufman et al., 2003), on board Aqua and Terra satellites, is used for various purposes: 1) to quantify and monitor the smoke AOD at the global scale, 2) to confirm the fires position and active period, and 3) to parameterize the smoke emission in the dispersion model. For the AOD we use the near real-time value-added MODIS AOD level 3 gridded product (MCDAODHD) based on MODIS level 2 aerosol products combined from Aqua and Terra satellites. The sensor resolution is 0.5º, imagery resolution is 2 km, and the temporal resolution is daily. For the fire information (position and active period), MODIS Fire and Thermal Anomalies products, either from Terra (MOD14), Aqua (MYD14) or a combination of them (MCD14), are used. Each MODIS active fire location represents the center of a 1-km pixel that is flagged by the algorithm as containing one or more fires within the pixel.

**2.3 Particle dispersion modeling**

**2.3.1 Model overview**

Back-trajectory and dispersion calculations are performed with the Hybrid Single Particle Lagrangian Integrated Trajectory Model (HYSPLIT; Stein et al., 2015; Rolph et al., 2017). HYSPLIT is developed at NOAA's Air Resources Laboratory and is one of the most widely used models for atmospheric trajectory and dispersion calculations. It is a complete system for computing simple air parcel trajectories as well as complex transport, dispersion, chemical transformation, and deposition simulations. The model calculation method is a hybrid between the Lagrangian approach and the Eulerian methodology. Apart from calculating back-trajectories, HYSPLIT is mostly used in this study to calculate the transport, dispersion, and deposition of emitted CO (used as a tracer of the transport) and particulate matter (black carbon, BC, and organic carbon, OC). The specificity of our HYSPLIT runs is that the heat release from the fires is used to estimate the smoke release height, i.e. no release heights were *a priori* set. The initial particle height is assumed equal to the final buoyant rise height as computed using the method of Briggs (1969) with the fire heat release given in input, implying that the final rise is a function of the estimated fire heat release rate, the atmospheric stability, and the wind speed. Stein et al. (2009) tested the sensitivity of HYSPLIT to fixed and variable release heights by comparing PM2.5 levels modelled and measured at the surface of northwestern US fires in September 2006. They found that the case when the heat release from the fire was used to estimate the release height showed the best performance, although they also concluded that the model is highly sensitive to variations in the smoke release height and to whether the smoke injection actually occurred below or above the planetary boundary layer. Rolph et al. (2009) also used HYSPLIT plume rise computation from the fire heat release.

For each fire simulated, a series of common parameters (in brackets we indicate HYSPLIT denomination) are necessary: location (Release location), start time (Release start time), duration (Release duration) and heat release (Heat release for plume rise). And for each chemical compound simulated (gas or particle), the emission rate (Emission rate) is necessary. Such information is extracted from the biomass burning emission estimates from GFAS (Global Fire Assimilation System; Kaiser et al., 2012) data from CAMS (Copernicus Atmosphere Monitoring Service). GFAS assimilates fire radiative power (FRP) observations from satellite-based sensors (Freeborn et al., 2014), namely MODIS/Aqua and Terra and SEVIRI (Spinning Enhanced Visible and InfraRed Imager), to produce daily estimates of biomass burning emissions. GFAS data (in brackets we indicate GFAS denomination) used in our work include daily information of the fire location and heat release (Wildfire radiative power), and for each chemical compound the emission rate (Wildfire flux). Data are available globally on a regular latitude-longitude grid with horizontal resolution of 0.125º x 0.125º. We used the current version of GFAS, i.e. GFAS v1.2. This work contains modified Copernicus Atmosphere Monitoring Service Information (CAMS, 2018).

**Table 1:** Instruments used in this study. The nomenclature $3\beta+2\alpha+1\delta$ stands for 3 elastic channels (here, 355, 532, 1064 nm), 2 Raman channels and one depolarization channel; $1\beta+1\delta$ stands for 1 elastic channel (here, 532 nm) and one depolarization channel; $3\beta$ stands for 3 elastic channels (here, 355, 532, 1064 nm). 8-$\lambda$ refers to the number (8) of wavelengths of the photometers.

| **Ground-based** | | | |
|---|---|---|---|
| | Station / Network | Latitude, longitude, altitude | Instrument type |
| Active | EV / EARLINET | 38.57N, 7.91W, 293 m asl | $3\beta+2\alpha+1\delta$ lidar |
| | AR / MPLNET | 37.10N, 6.73W, 59 m asl | $1\beta+1\delta$ lidar |
| | GR / EARLINET | 37.16N, 3.61W, 680 m asl | $3\beta+2\alpha+1\delta$ lidar |
| | MA / EARLINET | 40.45N, 3.72W, 669 m asl | $3\beta$ lidar |
| | BA / MPLNET | 41.39N, 2.11E, 115 m asl | $1\beta+1\delta$ lidar |
| Passive | Cerro Poyos / AERONET | 37.11N, 3.49W, 1830 m asl | 8-$\lambda$ sun-sky photometer |
| | Montsec / AERONET | 42.05N, 0.73E, 1574 m asl | 8-$\lambda$ sun-sky photometer |
| **Spaceborne** | | | |
| | Instrument | Satellite | Instrument type |

| Active | CALIOP | CALIPSO | 2β+1δ lidar |
| Passive | MODIS | Aqua and Terra | Moderate resolution imaging radiometer |
| | AIRS | Aqua | High-spectral resolution, multispectral infrared sounder |

The quantification of the HYSPLIT dispersion model uncertainties is not straightforward and it is usually performed through complex sensitivity studies (Mosca et al., 1998; Pielke and Uliasz, 1998; Straume, 2001; Warner et al., 2002). In general, the performance of dispersion models is largely attributed to uncertainty in the input fields (Challa et al., 2008). For our case, the GFAS data used for estimating the magnitude and timing of fire emissions have a typical uncertainty around 30% (Andela et al., 2013).

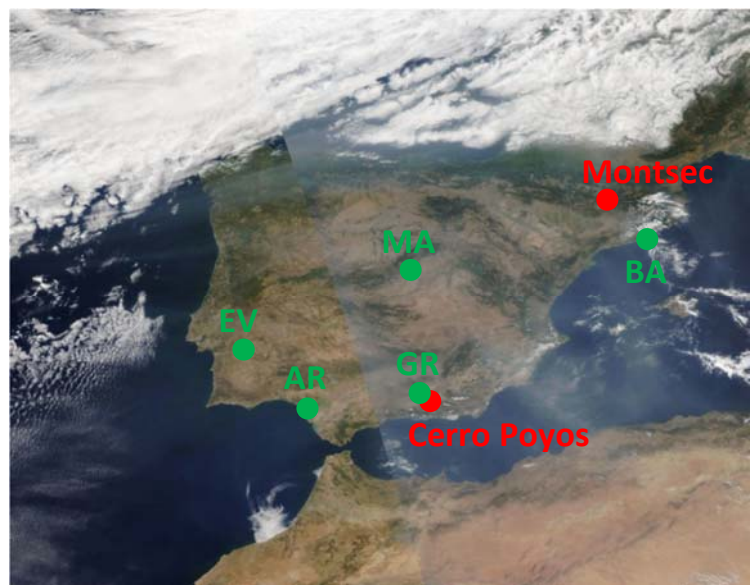

**Figure 1:** MODIS/Aqua corrected reflectance (true color) map centered over Spain on 8 September. Green bullets indicate lidar stations (EV: Évora, AR: El Arenosillo/Huelva, GR: Granada, MA: Madrid, BA: Barcelona) and red bullets indicate AERONET sites. Map created from https://firms.modaps.eosdis.nasa.gov/map/.

**2.3.2 Model parametrization**

The dispersion of CO, BC and OC is simulated in the forward direction, with a time resolution of 6 hours and at 15 altitude levels: one between 0 and 2.5 km and then 14 adjacent 1-km thick layers up to 16.5 km. The meteorology is taken from GDAS (Global Data Assimilation System) data with a horizontal resolution of 0.5º x 0.5º. Noteworthy is the fact that the first simulations with GDAS 1º x 1º meteorological data (not shown) simulated the dispersion of the smoke plume too far north reaching France and Germany, instead of the IP. The use of the finer resolution of 0.5º x 0.5º improved significantly the arrival location of the plume, and put forward the importance of the horizontal resolution of the meteorological data upon the correct dispersion of the emitted plume studied. The vertical limit of the internal meteorological grid of HYSPLIT was set to 20 km, which, according to the following sections, is well above the maximum height at which the smoke particles were observed. The daily number of active fires, the FRP per fire and the emission rate per fire and chemical compound are from GFAS 0.125º x 0.125º data. Since the time resolution of GFAS data is daily, the emission rate and FRP are assumed constant during the day the fires are active. In all simulations 2500 particles were released to calculate the transport.

In the case of CO, dry deposition is neglected and wet removal is parameterized with a Henry's law constant of $9.9 \times 10^{-4}$ mol atm$^{-1}$. BC (OC) is parameterized with the following values (Chin et al., 2002):

- Particle radius: 0.0118 (0.0212) μm.
- Particle density: 1.0 (1.8) g cm$^{-3}$.
- Gravitational settling velocity (for dry deposition): 0.5 cm s$^{-1}$ for both types.
- Scavenging coefficient in- and below-cloud (for wet deposition): $8 \times 10^{-5}$ s$^{-1}$ for both types.

**3 Methodology**

The proposed methodology is a two-way process, posterior to an initial phase (step 0) consisting in visualizing the "big picture" of the event at global scale with satellite images and back-trajectories. A flowchart of the methodology is shown in Figure 2. The first step of the methodology (step 1) consists in monitoring the smoke optical properties observed over the IP and their backward evolution back to the source with CALIOP retrievals. The second step (step 2) consists in parameterizing the smoke emission and run HYSPLIT forward simulations to obtain 4D (space and time) dispersion maps of the concentration of smoke-related compounds such as carbon monoxide, black carbon and organic carbon. The main contribution of the modelling in Section 6 is to support the possible hypothesis made along the discussion in Section 5.2.

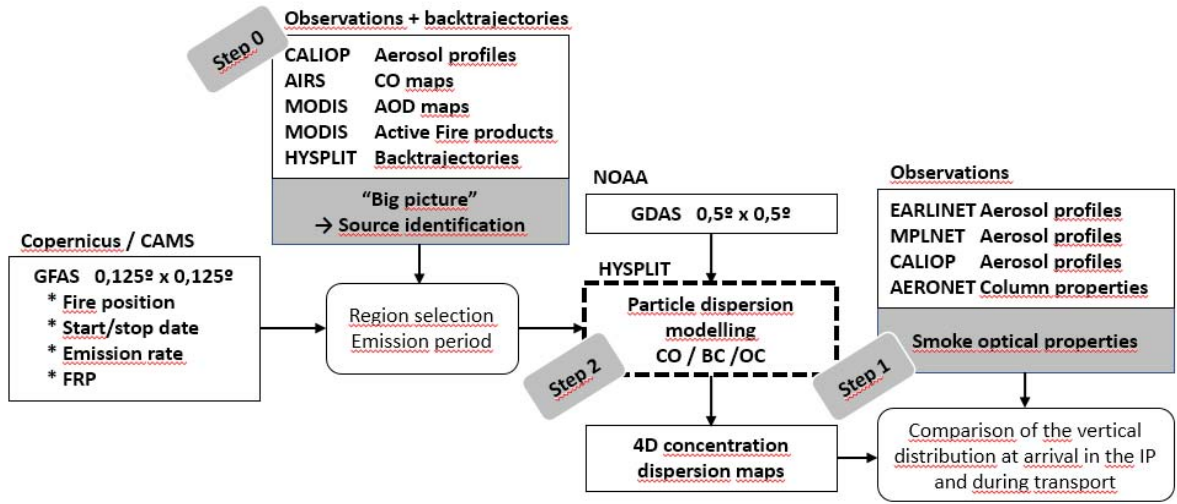

**Figure 2:** Flowchart of the methodology.

## 4 Canadian/United States fires and general overview

The first hint of the arrival and the presence of the smoke plume over the IP is given by the temporal evolution of a combination of AERONET parameters in Montsec and Cerro Poyos, namely the AOD at 440 nm, $AE_{440-870}$, and *FMF* (Figure 3). According to Sola et al. (2014) the mean AOD at 500 nm ($AE_{440-870}$) in Montsec during the month of September is ~0.1 (~1.5) which corresponds to an AOD at 440 nm of 0.12. In Montsec $AOD_{440}$ starts to exceed this value on 4 September, day from which the AOD increases continuously until it reaches its peak value of 0.55 (0.54) on 7 (8) September. These peak values of AOD are associated with values of $AE_{440-870}$ of 1.7 (1.6) on 7 (8) September. On both days the fine mode fraction is higher than 0.98, leaving basically no room for the presence of coarse mode smoke particles (diameter > 1μm). In Cerro Poyos, the background AOD at 440 nm is even lower than in Montsec, being smaller than 0.1 (AERONET, 2018). This value is exceeded from 5 September on, and the AOD increases until 7 September when it reaches its peak value of 0.62 with an associated $AE_{440-870}$ of 1.6. The fine mode fraction is higher than 0.88 on both 7 and 8 September. The main difference between Montsec and Cerro Poyos is their proximity to anthropogenic emissions: while Montsec is a remote site, far away from any industrial or large metropolitan area, Cerro Poyos, although higher in altitude, is only 12 km SE of the city of Granada (~600,000 inhabitants including metropolitan area). Due to its position with respect to Granada and the prevailing winds during the period under study, Cerro Poyos was downwind of the city. This has several implications: the AOD in Cerro Poyos shows a diurnal cycle related to the anthropogenic emissions of Granada, and the fine mode fraction is lower than in Montsec due to the same emissions. However one can appreciate from Figure 3 that during the night of 7-8 September *FMF* in Cerro Poyos is nearly 1 like in Montsec. In terms of AOD, the biomass burning contribution at both sites is roughly five times higher than the background values, and the peak values (~0.6) are considered very large for biomass burning long-range transport. In comparison the North American biomass burning event detected 15 days earlier (~22 August 2017) in northern Europe produced AODs at 500 nm near 1, $AE_{440-870} \sim 1.1$ and $FMF \sim 1$ (Ansmann et al., 2018).

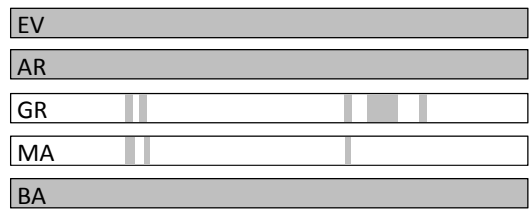

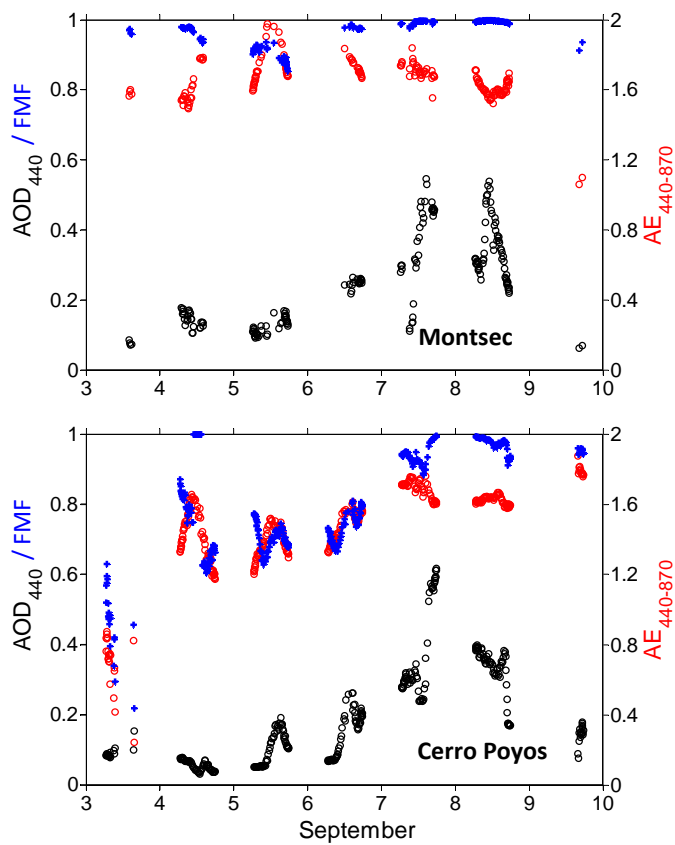

**Figure 3:** AOD$_{440}$ (black), FMF (blue) and AE$_{440-870}$ (red) in (top) Montsec, northeastern Spain, and (bottom) Cerro Poyos, south Spain. The gray areas in the bars on top of the figures indicate coincident lidar measurements.

To track back the plume transport in the atmosphere from the source to the IP, we use maps of columnar CO (AIRS) and AOD (MODIS; Figure 4) as well as CALIOP curtains and HYSPLIT back-trajectories (Figure 5). The combined day/night columnar CO maps are reported for the period 30 August – 8 September and a threshold of 95 ppb was applied in order to highlight strong concentrations. The active fires are indicated by a red star (Figure 4) centered in a region defined by the orange square visible in Figure 5 (bottom plot). This square covers the provinces of British Columbia and Alberta (Canada) and the states of Washington, Oregon, Idaho, Montana and northern California (US) where more than 90% of the active fires in North America are present during the period considered. The habitat type, a little more south than the Canadian boreal forests, corresponds to temperate coniferous forests (Ricketts et al., 1999). Some important forests in this region of North America are the National Forests of Wenatchee, Flathead, Nez Perce-Clearwater or Payette, among others, which, under the influence of both continental and maritime climates, produce a large variety of ecosystems ranging from wet, western redcedar bottoms to high alpine peaks, and forests of alpine larch and whitebark pine. So, from this region, a first plume (Plume 1) is released from the source region on 30 August, travels E-NE on 31 August and then eastwards on 1 and 2 September. On 3 September a second plume (Plume 2) is released from the same source region and starts travelling east, slightly SE. On 4, 5, and 6 September Plume 2 is carried by the jet stream and travels rapidly towards the east, while at the same time Plume 1 travels slowly eastwards above the Atlantic. The column concentration of Plume 2 is stronger than the one of Plume 1. On 7 September both plumes merge into one and reach the IP. The high MODIS AOD values on 8 September over the IP confirm that the high level of column CO is accompanied with high aerosol loads. These aerosols are also clearly visible especially in the eastern part of the IP as a gray/brownish smoke shroud on MODIS true color image of 8 September (Figure 1). The 10-day back-trajectories at selected heights are shown in Figure 5. Although we computed back-trajectories at all lidar stations, only Madrid is shown as point of arrival for the sake of clarity of the figure and because Madrid is located in the center of the IP. The selected heights (3, 6 and 11 km asl) have been chosen by looking at the smoke vertical distribution from the lidar data (see next section). All three trajectories pass over the region containing most of the active fires (orange square): the trajectory arriving in Madrid at 3 km passes over this region on 31 August at 08UT (7-8 days of transport) and the trajectories arriving at 6 and 11 km, very similar in path and speed, pass over the region of the fires on 3 September between 17 and 22UT (4-5 days of transport). These results suggest that the airmasses arriving above the IP at 3 km on 8 September picked up smoke most likely from Plume 1, while those arriving at 6 and 11 km most likely from Plume 2. During the transport CALIPSO orbits intersect the back-trajectories in space and time in two occasions: once southeast of Greenland on 4 September at 05:10 UT (D-4, 4 days before arrival in Madrid, intersects with Plume 1 which is 5 days old) at 3 km height, and another time on 3 September at 09:23 UT (D-5, intersects with Plume 2 which is less than 1 day old) at 6 and 11 km heights. On D-5 the shortest distance between CALIPSO curtain and the center of the region of the fires (orange square) is 700 km. On both occasions the cloud-free CALIOP curtains show clearly the large spatial extension of the smoke: 1700 km (below orbit) x 15 km (height) on D-4 and 1100 km (below orbit) x 8-9 km (height) on D-5. The attenuated backscatter of CALIOP on D-5 is clearly much stronger than on D-4 because of the proximity of the orbit to the source region. On D-5, in the southernmost part of the plume, most of the smoke between 4 and 7 km height is optically so thick that it attenuates the lidar signal below it.

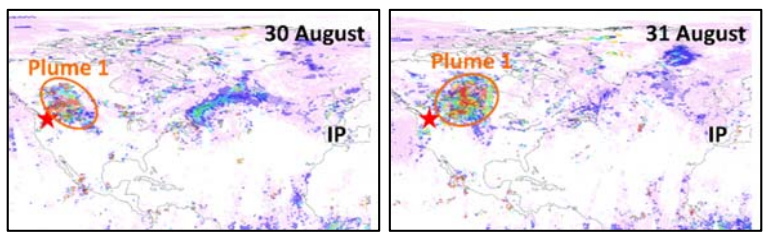

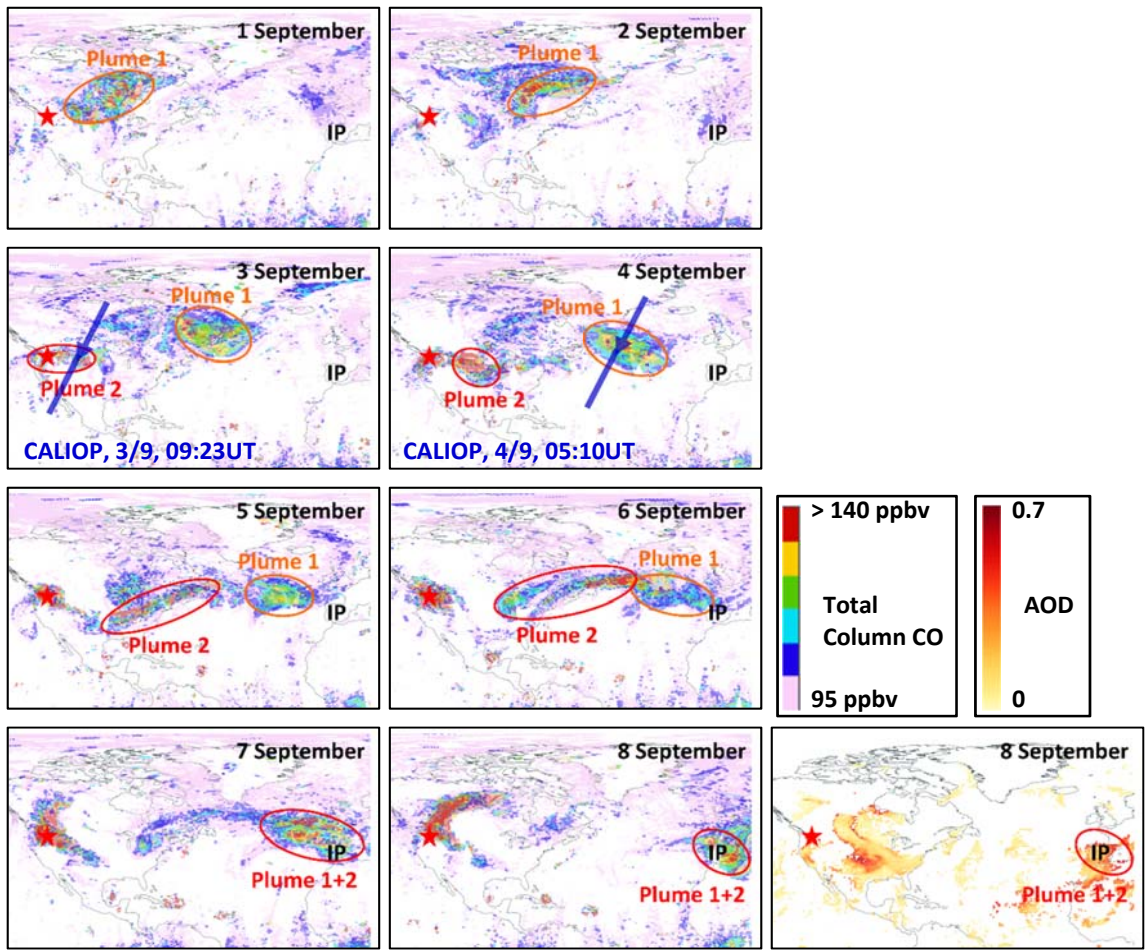

**Figure 4:** Total column carbon monoxide (day/night) from AIRS/AQUA from 30 August until 8 September. The extra plot at the bottom to the right represents the MODIS combined (Aqua and Terra) value-added AOD at 550 nm on 8 September. The red star indicates the position of the active fires. On the plots of 3 and 4 September the descending, nighttime orbits of CALIPSO are reported. Maps created from https://worldview.earthdata.nasa.gov/.

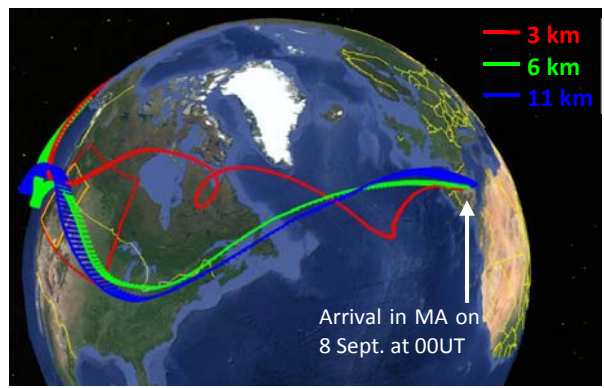

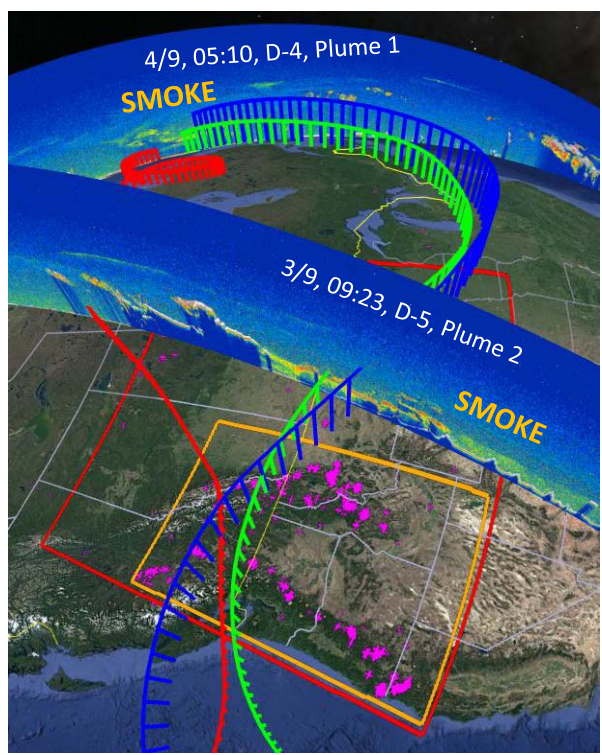

**Figure 5:** (top) 10-day back-trajectories, 1-hour resolution, arriving in Madrid, in the center of Spain, on 8 September at 00UT at heights of 3 (red), 6 (green) and 11 (blue) km; (bottom) Same back-trajectories, different viewing angle and superposition of CALIOP curtains on 4 September at 05:10UT (D-4, day-4 before arrival) and on 3 September at 09:23UT (D-5) where the smoke plumes, clearly visible, match very well in space and time with the back-trajectories. Pink crosses indicate active fires in the period 30 August – 5 September. The red rectangle of corner coordinates (125W, 40N; 93W, 58N) is the area in which the fires were taken into account in the dispersion modelling analysis (see Section 6). The orange rectangle simply highlights the region containing most of the fires. Maps created with Google Earth.

## 5 Optical properties of the smoke particles

Many papers, most of them listed in the literature overview of Ortiz-Amezcua et al. (2017) or of Haarig et al. (2018), deal with the optical properties of long-range transport smoke particles derived from observations of photometers, lidars or a combination of them. More general aerosol-typing literature based on lidar remote sensing and including biomass burning are available in Burton et al. (2012), Groß et al. (2013), Illingworth et al. (2015) and Baars et al. (2016; 2017).

### 5.1 Column-averaged properties

Figure 6 shows the spectral AAOD, SSA and asymmetry factor retrieved from AERONET sun-sky photometer measurements at Montsec and Cerro Poyos on 7 and 8 September. Several aspects are noteworthy. AAOD, similar at both sites in absolute values, is surprisingly very low. Compared to the climatological AAOD representative of boreal forests from US and Canada (Russell et al., 2010), recalculated from Dubovik et al. (2002), our AAOD values are 2 to 3 times lower. As a consequence of the small AAOD observed in Montsec and Cerro Poyos, SSA is large (~0.98 at 440 nm) and indeed much larger than the climatological values for boreal forest biomass burning (0.94 at 440 nm) from Dubovik et al. (2002). However it is in the range of values of SSA at 355 nm obtained in Europe by Markowicz et al. (2016), 0.91 – 0.99, and Ortiz-Amezcua et al. (2017), 0.965 – 0.991, in smoke plumes originating from North America in July 2013. In particular Markowicz et al. (2016) attribute these high SSA values to "a transformation of [biomass burning] during long-range transport […] and mixing of the [biomass burning] with non-absorbing aerosol species". The high transport altitude of the fire smoke observed over the IP in summer 2017 makes the second hypothesis (mixing with non-absorbing aerosol species) highly improbable. If these low AAOD and high

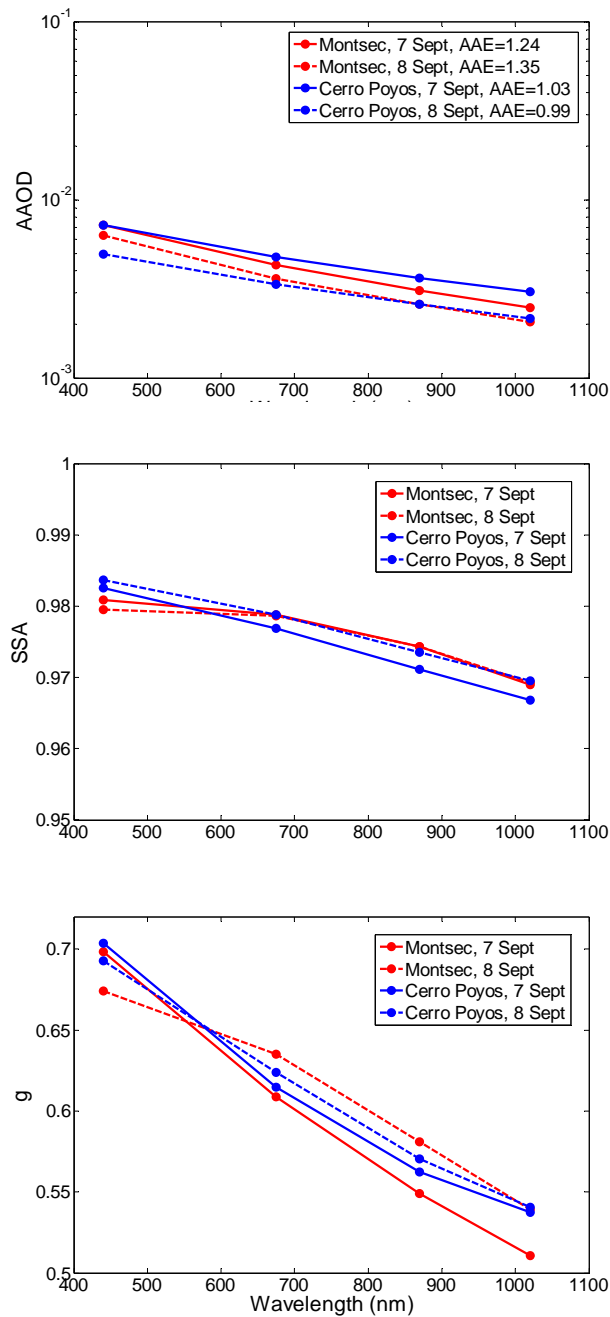

**Figure 6:** AERONET daily mean spectral (top) AAOD, (center) SSA, and (bottom) asymmetry factor at Montsec and Cerro Poyos on 7 and 8 September.

SSA were due to low BC emission at the source, the following rationale can be made. According to Radke et al. (1991) and more recently to Russell et al. (2014) the absorption properties of biomass burning in its smoldering combustion phase are lower than during its flaming phase, the reason being a larger production of black carbon in the flaming phase relative to the smoldering phase (Radke et al., 1991). In addition, smoldering combustion occurs over a much longer period of time relative to the comparatively short lives flaming phase of tree-crown fires of, e.g., pines, cedar or cypress that commonly populate temperate coniferous forests. These results suggest that the smoke particles observed over the IP might be the product essentially of smoldering combustion at the source. We also recall that level 1.5 AERONET data are not totally quality assured and that the values of AAOD and SSA should be taken with certain caution. The values and spectral behavior of $g$ in Montsec and Cerro Poyos are in good agreement with results for biomass burning aerosols from other studies (Dubovik et al., 2002; Sayer et al., 2014; Nikonovas et al., 2015), with $g$ presenting a sharp decrease with increasing wavelength. Nikonovas et al. (2015) distinguished the behaviour of fresh (within the first 24 h) and aged (more than 72 h) smoke and values reported therein for aged smoke agree quite well with the mean values obtained at Montsec and Cerros Poyos on 7 and 8 September 2017 (~0.70 at 440 nm and ~0.53 at 1020 nm). This spectral behavior is typical of the dominance of fine particles that are scatterers of solar radiation more efficient at lower wavelengths, with the forward scattering decreasing with increasing wavelength.

We analyze the wavelength dependence of AAOD by means of the absorption Ångström exponent, AAE, calculated between the wavelength of 440 and 870 nm. Although a clear difference is observed between both sites in terms of AAE: 1.24 < AAE < 1.35 in Montsec and 0.99 < AAE < 1.03 in Cerro Poyos, conclusions are not straightforward. We rely our discussion on the results of Lack and Cappa (2010). According to these authors, the AAE for pure BC cores varies in the range [-0.2, +1.3], for BC cores coated in non-absorbing matter (i.e. coated with a purely scattering shell) it can be as high as 1.6 -1.7, and for BC cores coated in absorbing matter, namely brown carbon (mildly absorbing organic matter; Andrea and Gelencsér, 2006), it is usually greater than unity even if for certain combinations of core/shell size pairs and values of the imaginary part of the refractive index, it can be close to unity. Thus the absolute attribution of BC or brown carbon is hampered when AAE < 1.6. However Lack and Cappa (2010) also showed that high SSA values (> 0.9) could only be achieved for BC cores coated in absorbing matter. The results allow us to conclude (i) that, without any doubt, the AAE values in Montsec (~ 1.3) are representative of brown carbon (or BC coated in brown carbon), likely contained in the long-range transport smoke plume detected, and (ii) that the AAE ~ 1 in Cerros Poyos is most probably caused by brown carbon from biomass burning origin and maybe pure BC from the anthropogenic fossil fuel emissions of the nearby city of Granada. Another possible reason for AAE ~ 1 in Cerros Poyos may be the presence of nearby persistent local fires in Sierra Morena, approximately 150 km northwest of Cerro Poyos. For comparison, Bergstrom et al. (2007) measured an AAE of 1.45 in the range 325 – 1000 nm in a plume of South Africa biomass burning with data from the SAFARI (Southern Africa Regional Science Initiative) campaign.

**5.2 Vertically-resolved properties**

To relate smoke optical properties and their vertical distribution, we use ground- and space-borne lidar profiles. The availability of lidar measurements in the period 3 – 9 September is indicated by the gray areas in the bars of Figure 3. Because of the high aerosol load and the high vertical extension of the plumes (> 10 km) on the night of 7 to 8 September and their implication on the signal-to-noise ratio of the lidar signals and thus on the quality of the inversions, Raman inversions were performed only the night of 6 to 7 September. In Figure 7 we show the result of a Raman inversion in Évora on 7 September between 04 and 06UT. Although this measurement time is a few hours before the arrival time fixed for the back-trajectory simulations (8 September at 00UT) our back-trajectory analysis (not shown) confirms that air mass paths were very similar during the 48 hours of both days 7 and 8 September. A series of quality checks have been applied to Évora lidar profiles: negative optical properties are not considered and intensive properties ($\alpha - AE$, $\beta - AE$, $CR$, $LR$ and $\delta^p$, see caption of Figure 7 for symbol definition) are calculated only for optical properties greater than a minimum threshold in order to guarantee the presence of aerosols and to avoid physically meaningless retrievals. In addition to the profiles, smoke layer-mean values are given in two altitude ranges corresponding to the mid and upper troposphere. These layer mean values are also reported in Table 2. We find smoke particles up to 12.7 km, below the tropopause height, with a clear plume extending from 2.3 to 8.1 km and a very shallow one from 11.5 to 12.7 km. The AOD at 532 nm is 0.24. The color ratio is significantly different in the two altitude levels considered (~2.49 in the mid troposphere and ~3.31 in the upper troposphere). For comparison Haarig et al. (2018) found color ratios of 1.8 and 2.3 in the troposphere and the stratosphere, respectively, for the North American biomass burning detected in northern Europe 15 days earlier (22 August). Our higher values indicate particles of smaller size. This finding is corroborated by the columnar effective fine mode radius measured by AERONET in Montsec and Cerro Poyos on 7 September which vary in the range 0.14 – 0.18 μm, while values larger than 0.23 μm were found during the 22 August event (Ansmann et al., 2018). $\alpha - AE$, only retrieved in the mid troposphere, is 1.51. For comparison Haarig et al. (2018) found a $\alpha - AE$ of 0.9 in the troposphere. The difference with our $\alpha - AE$ is probably due to different absorption properties: low in our case and rather large on 22 August (Ansmann et al., 2018). Low absorption properties yield to $\alpha - AE$ very similar to the scattering Ångström exponent. According to Valenzuela et al. (2015) scattering AE larger than 1.5 indicates that submicron particles dominate the aerosol size distribution, which is in agreement with our findings.

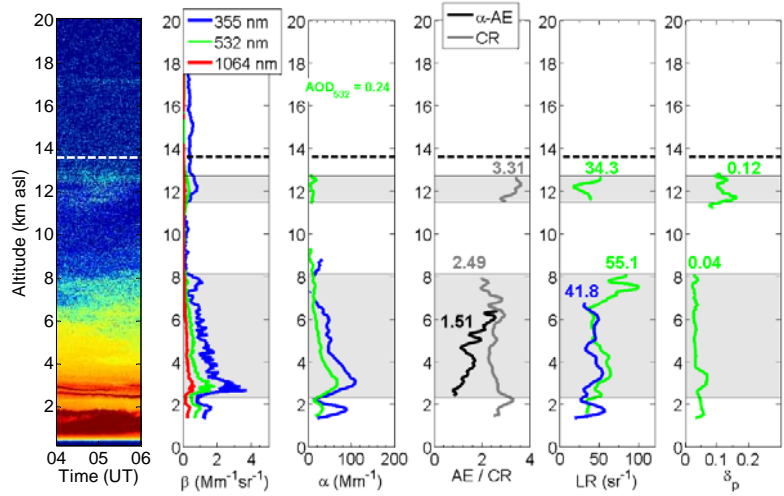

**Figure 7:** Nighttime multi-wavelength lidar inversion in Évora on 7 September between 04 and 06UT. The first plot represents the quicklook of range-square corrected signal at 1064 nm in arbitrary units. $\beta$ is the particle backscatter coefficient, $\alpha$ the particle extinction coefficient, $\alpha-AE$ the extinction-related AE, $CR$ the color ratio, $LR$ the lidar ratio and $\delta_p$ the particle depolarization ratio. Mean values in the mid troposphere and stratosphere (as depicted by the gray rectangles) for $\alpha-AE$, $CR$, $LR$ and $\delta_p$ are reported in the plots. The horizontal dash lines at 13.6 km indicate the tropopause height calculated with 1° x 1° GDAS data.

The lidar ratio at 532 nm is 55.1±14.2 and 34.3±10.5 sr in the mid and upper troposphere, respectively. Our $LR_{532}$ values are slightly lower than those of Haarig et al. (2018), 65 – 80 sr, if we take into account the standard deviations associated to our retrievals, but are definitely in the range of literature values (26 – 80 sr) for North American biomass burning detected in Europe (Ortiz-Amezcua et al., 2017). A lower lidar ratio in the upper troposphere compared to the mid troposphere might indicate less absorbing particles in higher altitude (Ortiz-Amezcua et al., 2017), maybe related to a lesser amount of BC with respect to organic carbon, or to a lesser degree of coating on BC since coatings on BC enhance scattering and absorption properties (Cheng et al., 2014). These hypotheses are further investigated in the next section. At 355 nm we find in Évora $LR_{355} = 41.8 \pm 6.8\ sr$ in the mid troposphere, which is in good agreement with the range of 40 – 45 sr found by Haarig et al. (2018) and with the range of literature values (26 – 80 sr) from Ortiz-Amezcua et al. (2017).

**Table 2:** Layer mean values of the color ratio, the Ångström exponent, the lidar ratios at 355 and 532 nm and the particle depolarization ratio at 532 nm in Évora, and from CALIOP on D-4 and D-5.

| Parameter | UTLS | | | Mid troposphere | | |
|---|---|---|---|---|---|---|
| | Évora 7/9 at 04UT | CALIOP D-4 Plume 1 | CALIOP D-5 Plume 2 | Évora 7/9 at 04UT | CALIOP D-4 Plume 1 | CALIOP D-5 Plume 2 |
| Color ratio | 3.31±0.27 | 3.06±1.19 | - | 2.49±0.24 | 2.17±0.28 | 1.86±0.36 |
| α-AE | - | NA | NA | 1.51±0.76 | NA | NA |
| $LR_{355}$ (sr) | - | NA | NA | 41.8±6.8 | NA | NA |
| $LR_{532}$ (sr) | 34.3±10.5 | NA | NA | 55.1±14.2 | NA | NA |
| $\delta_p$ | 0.12±0.02 | 0.12±0.03 | - | 0.04±0.01 | 0.05±0.01 | 0.05±0.02 |

Last but not least, the analysis of the profile of the particle depolarization ratio at 532 nm also reveals interesting results. The layer mean values of $\delta_p$ are 0.04 and 0.12 in the mid- and upper troposphere, respectively. While the mid troposphere value falls in the range of literature values (Ortiz-Amezcua et al., 2017) and indicate spherical or almost spherical smoke particles, the value of 0.12 in the upper troposphere is rather unusual. Some works investigating the inter-continental transport of North American fire smoke to Europe from August 2017 also report unusually high depolarization ratios (Khaykin et al., 2018; Haarig et al., 2018; Hu et al., 2018; Sicard et al., 2018) up to 0.20 at 532 nm in the stratosphere. The causes of such high depolarizing capabilities of smoke particles are still not well understood. Recently Burton et al. (2015) made a nice discussion based on literature to explain the high values of three-wavelength depolarization ratios and their spectral dependence that they observed for smoke particles from North American fires retrieved by high-spectral resolution lidar. They proposed two possible explanations of the depolarization by smoke: the "lifting and entrainment of surface soil into the smoke plume and asymmetry of smoke particles themselves". Haarig et al. (2018) hypothesized that high $\delta_p$ values may be the result of dried out smoke particles (relative humidity ~0 %) with a non-spherical shape. This hypothesis, however, is probably unlikely in the range of altitude considered here (< 13 km over the IP) as radiosoundings in Barcelona (not available in Évora) on 8 September at 00 UT indicate a relative humidity in the range 20 – 30 % in the upper troposphere. At this stage of the paper, our intention is not to give a single explanation of our high $\delta_p$ values, as we believe that the main features observed over the IP (injection in the upper troposphere, low absorption and high depolarization properties) are somehow connected, but to list some fire characteristics and physical/chemical mechanisms which could lead to such features: the burnt material at the source (BC and OC contents), flaming versus smoldering phases, fire power, BC aging processes (coagulation, condensation, and heterogeneous reactions) during transport resulting in changes in its morphology and mixing state, relative humidity. Literature on these issues can be found in Fromm et al. (2003; 2008), Zhang et al. (2008), Lack and Cappa (2010), Adachi et al. (2010), Cheng et al. (2014), China et al. (2015). Forrister et al. (2015), Burton et al. (2015), among many others, and will be used in the discussion of the next section. To analyze the spatio-temporal evolution of the smoke transport, we compare the smoke $CR$ and $\delta_p$ profiles from Évora (7 September at 04 UT) with CALIOP retrievals in Plume 1 (5 days old, D-4) and in Plume 2 (fresh < 1 day, D-5). CALIOP retrievals are shown in Figure 8. CALIOP quicklooks of the total attenuated backscatter at 532 nm show a spatial extension clearly larger for Plume 1 than for Plume 2 both horizontally and vertically. Plume 1 extends up to ~15 km and into the stratosphere while Plume 2 stays in the troposphere below 9 km. This result suggests that the UTLS injection of smoke particles does not occur immediately a few hours after fire ignition but during the transport. Indeed Cammas et al. (2009) simulated with the anelastic non-hydrostatic mesoscale model Meso-NH the time needed for a boundary layer tracer to reach the tropopause to be about 7.5 hours. Logically, the particle backscatter coefficient at 532 nm is much stronger in Plume 2 (> 5 Mm$^{-1}$sr$^{-1}$ below 4 km at latitudes of 48 – 50ºN) resulting in a high AOD at 532 nm of 1.20 (versus 0.78 for Plume 1). The color ratios in the troposphere are 2.17 and 1.86 for Plume 1 (5 days old) and Plume 2 (< 1 day), respectively, indicating a decrease of the particle size as the plume gets older. In Évora $CR$ is 2.49. In the stratosphere the color ratio of Plume 1 is 3.06, while it is 3.31 in the upper troposphere in Évora. Given the large standard deviation of CALIOP $CR$ retrieval in the stratosphere (Table 2), the relatively small difference between both values (3.06 and 3.31) cannot be interpreted as a decrease of particle size. In fact, once in the UTLS the smallest particles (with radii < 0.5 μm), tend to maintain at their altitude level or to ascend. Rohatschek (1996) and Pueschel et al. (2000) explained the self-lofting of UTLS-level BC with the gravito-photophoresis mechanism consisting in sunlight-induced upward forcing. It is interesting to note a significant increase of $CR$ (> 3) close to the ground in Plume 2 which probably reflects freshly emitted, small soot particles, before they undergo any of the various aging processes that lead to their size increase. The CALIOP particle depolarization ratio at 532 nm in the troposphere is 0.05 in both plumes, a value similar to $\delta_p = 0.04$ found in Évora in the mid troposphere. It is an indication that the smoke particle depolarizing capabilities, and subsequently also their shape, in the troposphere are stable during transport. In the stratosphere $\delta_p$ in Plume 1 increases from 0.04 to a peak value of 0.16, the mean value being 0.12. In Évora the same value of 0.12 is found in the upper troposphere between 11.5 and 12.7 km. As far as Plume 1 is concerned, the smoke particles reached the UTLS in less than 5 days after their release in the atmosphere and it seems that the smoke particle depolarizing capabilities (and thus their shape) at UTLS level are also stable during transport. The quasi-linear increase of $\delta_p$ with height may be an indication of the height-dependence of the ongoing aging processes leading to the transformation of the smoke particle depolarization properties from low- (0.04 at 12.25 km) to moderately-depolarizing (0.16 at 14.95 km). In Évora $\delta_p$ in the upper troposphere does not seem height-dependent as the particles must have already undergone these aging processes.

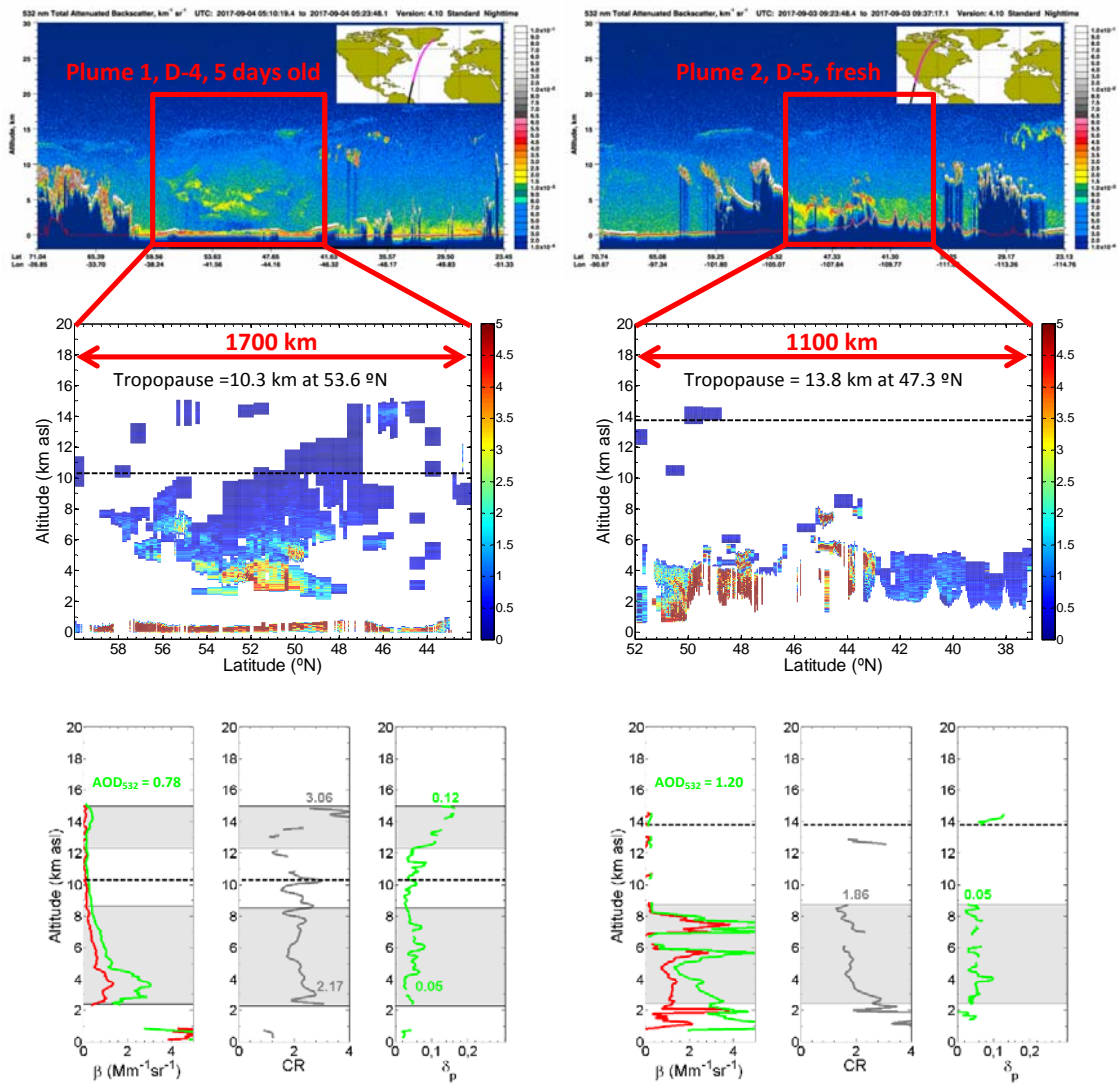

**Figure 8:** CALIOP images and products on (left) 4 September at 05:10UT (D-4, Plume 1 released 5 days earlier) and (right) 3 September at 09:23UT (D-5, Plume 2, fresh < 1 day). (top) CALIOP quicklooks of the total attenuated backscatter signal at 532 nm; (center) CALIOP quicklooks of the retrieved backscatter coefficient at 532 nm restricted to the smoke plume (red squares); (bottom) CALIOP mean profiles of backscatter coefficient at 532 and 1064 nm, the color ratio and the particle depolarization ratio at 532 nm. The horizontal black dash lines indicate the tropopause height calculated with 1º x 1º GDAS data.

To close this section, we compare the time-height evolution of mid and upper tropospheric particle depolarization ratio at all Iberian lidar stations (capable of measuring particle depolarization) plus CALIOP, and the dependency of $\delta_p$ versus $CR$ in Évora and Granada plus CALIOP. The results are shown in Figure 9. The reason for choosing to plot $\delta_p$ versus $CR$ is twofold: 1) they are the two intensive parameters provided by CALIOP, and 2) low-level aerosol typing is possible with these parameters (Groß et al., 2013), although the classification they propose also includes the lidar ratio. The profiles in the IP were selected during the night of 7 to 8 September, close to the back-trajectory arrival time (8 September at 00UT), according to measurement availability and clear-sky conditions. They all fall around the back-trajectory arrival time – 3/ + 1 hour. Contrarily to the observations of North American smoke in the stratosphere in France (Khaykin et al., 2018; Hu et al., 2018) and Germany (Ansmann et al., 2018; Haarig et al., 2018) earlier in August, 2017, over the IP no aerosols are observed in the stratosphere in the period considered. At all stations of the IP, a continuum of aerosols is observed up to the upper troposphere: aerosols are present in the whole troposphere. In order to identify representative layers and give layer mean values, we selected in both the mid and upper troposphere the layers centered around the backscatter coefficient peak value in each altitude range. This methodology guarantees a higher representativeness of the smoke particle properties, but in certain cases the selected layer may be spatially thin which may bias the interpretation of the top plot of Figure 9. Hence the layer height and its thickness represented in this plot has to be interpreted as the layer of maximum intensity, i.e. of maximum aerosol load. Before entering in the discussion, it is worth noting the important difference between the AOD at 532 nm in Barcelona (0.65) and the rest of the stations of the IP (0.27 – 0.34). This difference is indeed not that surprising if we look back at MODIS AOD on 8 September (Figure 4) which clearly shows a decreasing AOD tendency along the axis NE-SW. In the mid troposphere, with the exception of Granada, all measurements including CALIOP give a particle depolarization ratio of 0.05 – 0.06. This result reflects again that in the mid troposphere the smoke particle depolarization ratio was neither time- nor plume-dependent. The particle depolarization in Granada at ~7 km is 0.01 and clearly indicates non-depolarizing particles, slightly different from what is observed over the other stations of the IP in the mid troposphere. The layers of maximum intensity are rather high (between 6 and 8.3 km) and are higher in Évora, Granada and El Arenosillo/Huelva than in Barcelona. The center of the youngest plumes (CALIOP) are slightly below the layers of maximum intensity detected over the IP. In the upper troposphere $\delta_p$ varies between 0.09 and 0.10 over the IP, while in the low stratosphere $\delta_p = 0.12$ is found for Plume 1 (CALIOP, D-4). Given the large standard deviations associated to the retrieval of $\delta_p$ in the UTLS, these findings are here again not sufficient to point out a clear difference between the plume over the IP on D-0 and CALIOP on D-4. In case this difference is real, at this stage our findings only allow to give hypothetical explanations to be taken with care, listed in order of likelihood: 1) the stations over the IP are representative of Plume 1 + Plume 2 while CALIOP D-4 is representative of Plume 1; 2) on D-0 the plume is in the upper troposphere while on D-4 it is in the stratosphere; and 3) a transformation of the smoke depolarizing capabilities between D-4 and D-0.

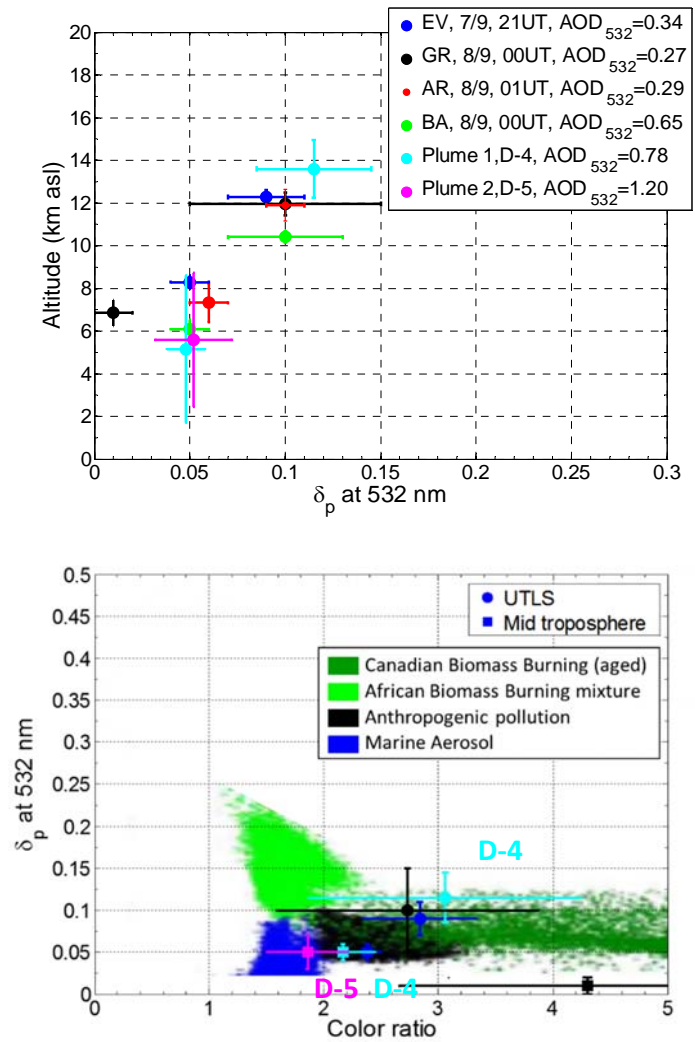

**Figure 9:** (top) Mid and upper tropospheric layer mean particle depolarization ratios at 532 nm at all Iberian lidar stations on the night of 7 to 8 September. Cyan and Purple bullets represent CALIOP measurements. The vertical bars indicate the vertical extension of the smoke layers of maximum intensity (base to top height). The horizontal bars indicate the standard deviation associated to $\delta_p$ in these layers. (bottom) Layer mean particle depolarization ratios at 532 nm vs. layer mean color ratio. The bullet color code is the same as in the top plot. We have reported four aerosols classes adapted from Groß et al. (2013). The vertical and horizontal bars indicate the standard deviation associated to $\delta_p$ and $CR$, respectively.

The bottom plot of Figure 9 represents pairs of ($\delta_p$, $CR$) mean layer values in Évora and Granada (D-0) and in Plume 1 (D-4) and Plume 2 (D-5). The color-coded shaded areas, representative of different aerosol classes, are adapted from Groß et al. (2013). The results can be summarized as follows:

- UTLS-level smoke particles have large color ratios (~2.5 – 3) and moderate particle depolarization ratios (~0.10). Between D-4 and D-0 a small decrease of both parameters, within the statistical variability of one another, is noted.

- In the mid troposphere, $\delta_p = 0.05$ is stable with time, except in Granada. CALIOP $CR$ values are smaller than in Évora (and also Madrid (not shown) as $CR \simeq 2.95$ in the smoke layer centered around 6 km on 7 September at 21UT; in Madrid no smoke layer was observed in the UTLS), indicating that as the smoke gets closer to its arrival in the IP, the particles get smaller. The difference between CALIOP $CR$ values cannot be evaluated since they correspond to two different smoke plumes which may initially have different morphology and thus different optical properties. The results obtained in Granada $\left(\delta_p = 0.01, CR = 4.30\right)$ are an indication of ultrafine, non-depolarizing particles and reveals a clear difference in the smoke properties with the rest of observations in the mid troposphere. The back-trajectories in all three southern stations (EV, AR and GR) are very similar and do not allow to give an explanation related with long-range transport. In turn, locally, Granada may have been exposed to nearby persistent fires in Sierra Morena, approximately 150 km northwest of the city. Fresh smoke produces low $\delta_p$ and large $CR$, but this is only an hypothesis at this stage.

- The pairs of ($\delta_p$, $CR$) fall in the Canadian biomass burning type, but often on the edges. In the mid troposphere, except in Granada, the pairs of ($\delta_p$, $CR$) actually overlap between the classes of Canadian and African biomass burning and marine aerosols. It is worth recalling that the fires studied are not exactly "Canadian biomass burning", which stands for boreal forest fires in the literature, but fires from temperate coniferous forests. This result calls for further investigation on biomass burning properties in relation to their origin which goes beyond the usual Amazonian, African and North American classes.

Note *en passant* that the mid and upper tropospheric values of $\delta_p$ and $CR$ in Évora on 7 September at 21UT (close in time to the back-trajectory arrival time of 8 September at 00UT, so called D-0) are not significantly different from the values found on 7 September at 04UT, for which the Raman inversion was performed (see Figure 7 and associated text).

# 6 UTLS injection and inter-continental transport

In order to investigate the role of each of the plumes identified in Figure 4, each plume is simulated separately and then together. The emission of Plume 1 is set to 30/8 – 1/9 (3 days) and the emission of Plume 2 to 2/9 – 5/9 (4 days). Only fires falling inside the red square defined in Figure 4 are considered. For each chemical compound (CO, BC and OC), three simulations are run, corresponding to:

- Plume 1 (noted P1 from now on) from 30/8 until 8/9 with emissions limited to the period 30/8 – 1/9.
- Plume 2 (P2) from 2/9 until 8/9 with emissions limited to the period 2/9 – 5/9.
- Plume 1 + Plume 2 (P1+2) from 30/8 until 8/9 with emissions from 30/8 until 5/9.

From the results of Section 4, the hypothesis is implicitly made that the emitted matter before 30/8 and after 5/9 is not affecting the IP on 7 and 8 September. The fire characteristics are summarized in Table 3.

Before entering in the discussion, we recall the questions raised in Section 5.2 and left opened: 1) injection mechanisms responsible of the injection in the upper troposphere, 2) smoke particles with low-absorbing properties and decrease of the absorption properties with height, 3) high depolarization properties, and 4) differences observed between the smoke plume observed in the IP (D-0) and the younger plume (D-4) observed by CALIOP.

**Table 3:** Characteristics of the fires at the origin of the emission of Plume 1 (emission: 30/8 – 1/9), Plume 2 (emission: 2/9 – 5/9) and for the whole period (emission: 30/8 – 5/9). The data are from GFAS daily estimates of biomass burning emissions.

|  |  | **P1** | **P2** | **P1+2** |
|---|---|---|---|---|
| Simulation period |  | 30/8 – 8/9 (10 days) | 2/9 – 8/9 (7 days) | 30/8 – 8/9 (10 days) |
| Emission period |  | 30/8 – 1/9 (3 days) | 2/9 – 5/9 (4 days) | 30/8 – 5/9 (7 days) |
| Number of active fires |  | 836 | 772 | 1073 |
| Number of active fires x day |  | 1843 | 2123 | 3966 |
| FRP per fire (MW) | Min | 0.1 | 0.1 | 0.1 |
|  | Mean | 95.1 | 137.0 | 117.5 |
|  | Max | 5405.7 | 7162.2 | 7162.2 |
| Number of fires with FRP > 50 MW |  | 232 | 190 | 277 |

|  |  | **P1** | | | **P2** | | | **P1+2** | | |
|---|---|---|---|---|---|---|---|---|---|---|
|  |  | **CO** | **BC** | **OC** | **CO** | **BC** | **OC** | **CO** | **BC** | **OC** |
| Emission rate per fire (T h$^{-1}$) | Min | 0 | 0 | 0 | 0 | 0 | 0 | 0 | 0 | 0 |
|  | Mean | 26.63 | 0.15 | 2.23 | 32.39 | 0.17 | 2.79 | 29.71 | 0.16 | 2.53 |
|  | Max | 2031.15 | 10.80 | 175.79 | 2561.98 | 13.63 | 221.73 | 2561.98 | 13.63 | 221.73 |

During the emission period 30 August – 5 September a total of 1073 fires were detected in the domain considered (red square, Figure 5), which in terms of fires per day, defined as the sum of all fires multiplied by the number of days they were active, represents nearly 4000 emitting fires. From these numbers one can deduce that, on average, each fire had an emission duration of approximately 4 days. The mean FRP is 117.5 MW and it is approximately 50 % higher during P2 than during P1. Maxima are also higher during P2 (7162.2 MW) than during P1 (5405.7 MW). Two hundred and seventy seven fires had a FRP larger than 50 MW. For comparison, Ansmann et al. (2018) reports a number of 10000 fires with FRP larger than 50 MW in Canada for the month of August 2017.

Before presenting the results of the dispersion analysis, we will make a point about the geographic location of the fires of this study. A notable difference between the fires of August (Khaykin et al., 2018; Ansmann et al., 2018; Haarig et al., 2018; Hu et al., 2018) vs. September 2017 (this study) is the latitude at which they occurred. In August the most intense fires were located in Canada, 49 < latitudes < 67 ºN (see Ansmann et al., 2018), while the emission region considered in our study goes from British Columbia down to northern California, 40 < latitudes < 53 ºN. In this lower part of the mid-latitude region, air masses can be under the influence of either the polar or the subtropical jet streams, and therefore be entrained either north- or south-ward, respectively. The latitude difference, 49 – 67 ºN vs. 40 – 53 ºN, also results in a higher tropopause height in September than during the August event, and also a thicker troposphere-stratosphere transition layer due to the vicinity of the subtropical jet in September (Pan et al., 2004). It has also another important implication: the material burnt, and consequently the content of emitted CO, BC and OC, are different. While the fires in August were from boreal forests, in September the fires occurred in a region of temperate coniferous forests. According to Lavoué et al. (2000) the main difference between boreal and temperate forests does not rely on the canopy itself, but in the shrubland and the grassland which are more abundant in temperate forests. McMeeking et al. (2009) who made controlled laboratory burns of Alaskan spruce and forest floor (duff), among other fuels, found that the forest floor has a strong contribution from smoldering combustion, but a lower carbon monoxide emission factor than most of the other fuels because it contains less carbon per mass unit. This result supports low BC and OC contents at the source and thus low absorption properties as discussed in Section 5.1.

Results are shown in Figure 10 and Supplements (S) 1-12. Figure 10 shows the dispersion maps of CO, BC and OC over the IP at time of arrival on 8 September at 00UT in terms of column density, i.e. the concentration integrated along the vertical axis. S1-S4, S5-S8 and S9-S12 are 6-hour time resolution animated gif images of the dispersion of, respectively, CO, BC and OC for the total column density, and the concentration at 3, 6 and 11 km. The heights of 6 and 11 km are representative of the mid- and upper troposphere, where smoke particles were detected in the IP. The images of the total column density (S1, S5 and S9) are the same as in Figure 10, i.e. a dispersion map on top and a longitudinal cross-section below at the latitude of Madrid, taken as a central point in the IP. On all dispersion maps, the color bar for P1 goes from red (low) to yellow (high) and for P2 from blue (low) to green (high). On a horizontal scale all three compounds have similar dispersion patterns. For this reason, the interpretation of the dispersion maps is made independently of the compound. Partly because the simulation time of P1 is longer relative to P2, its dilution in the northern hemisphere is wider and circulations around the globe start to be visible, although in low concentration level, in the final dispersion maps of all compounds. On 8 September CO/BC/OC from P1 are present around the pole and also in eastern Russia. Interpretations of S1, S5 and S9 confirm that:

- P1 is transported northeast-ward the first four days of the simulation. On 3 September a large swath of the US is covered by P1. This feature is confirmed by OMPS images (Seftor, 2017c). Later, as a large smoke tongue travels slowly over the Atlantic towards the IP (this transport coincides with the interpretation of the satellite images in Figure 4, see Section 4), large scale jets make the plume start meandering anti-clockwise around a point centered initially above Iceland which drifts slowly with time towards Ireland.

- P1 reaches the IP (Madrid) on 6 September at 12UT with column density levels of CO of 0.02 mg m$^{-2}$. For comparison on 8 September at 00UT the CO column density of P1 is on the order of 0.83 mg m$^{-2}$.

- P2 is travelling eastward since the first day of emission. On 4 September at 18UT one can already observe the beginning of the stretching of P2 located on the cyclonic-shear side of a strong jet, probably of subtropical origin since it ends up in northern Africa. Later a relatively thin smoke tongue travels rapidly along the large scale jet towards the IP. Residual smoke from P1 is also marginally carried with this flow.

- P2 reaches the IP (Madrid) on 7 September at 12UT with column density levels of CO below 1.20 mg m$^{-2}$. For comparison on 8 September at 00UT the CO column density of P2 is on the order of 18.60 mg m$^{-2}$. This result indicates that at the peak of the event the CO level observed over the IP and emitted by P2 is roughly 20 times larger than the one emitted by P1. For comparison, Yurganov et al. (2001) reports values of total column CO in Moscow, Russia, in the vicinity of strong wildfires ~50 times larger than the our values of ~20 mg m$^{-2}$.

As far as the vertical transport is concerned, we first analyze S3, S7 and S11 to identify where and when the injection at 6 km (mid troposphere) occurs, and then S4, S8 and S12 for the injection at 11 km (upper troposphere). CO/P1 (S3) appears for the first time at 6 km 18 hours after the first emission of the fires and close to the source region. CO/P2 appears for the first time at 6 km much later, ~30 hours after the emission, but also much farther, ~2000-3000 km from the source. In the upper troposphere, CO/P1 (S4) appears for the first time at 11 km 36 hours after the first emission of the fires and at ~2000 km from the source region, while CO/P2 appears for the first time at 11 km 60 hours after the emission and about 4000 km east of the source region. The time difference between injections from 6 to 11 km is 18 hours for P1 and 30 hours for P2, indicating a much faster ascending rate for P1 than for P2, despite higher FRPs during P2 relative to P1 (Table 3). It seems there is a tradeoff between vertical and horizontal transport: slow horizontal transport is favorable to vertical motion whereas strong horizontal transport reduces it. In the case of P2, it is highly probable that the strong jets leading to its fast transport towards the IP contributed at the same time to limit its vertical transport. In addition, we believe that the injection at higher altitudes of P1 is favored by its transport above the region of Lake Winnipeg (a large lake visible on the Supplements to the southwest of Hudson Bay) where wildfires are active, especially in the northern part of the lake. The hot region of the active fires is prone to increase the convection of upper air masses travelling above it. The analysis of S7 and S11 (BC and OC at 6 km) and S8 and S12 (same compounds at 11 km) indicates that the ascending rate of BC and OC is slower than for CO. BC and OC reach the altitude of 6 km approximately 6 hours after CO does, and they reach the altitude of 11 km approximately 12 hours later than CO. No significant difference is observed between the two types of particles in terms of ascending rate.

As far as the maximum injection height is concerned, interestingly none of the altitude levels is empty, indicating that the dispersion model injects smoke at all altitude levels considered, i.e. up to 16 km. However, above 10 km the number of pixels with non-zero concentration significantly decreases. We investigate the maximum injection height calculated by the model by defining a threshold of significant aerosol load at a given height when the probability of occurrence is greater than 2 %, i.e. when more than 2 % of the pixels at a given height are filled with non-zero values. With such a criterion, we find that CO/P1 (CO/P2) generally stays below 13 (9) km, BC/P1 (BC/P2) below 12 (9) km and OC/P1 (OC/P2) below 12 (11) km. These approximations of the maximum injection heights are in good agreement with the profiles of both the ground-based lidar stations and CALIOP. One sees clearly that the injection at high levels is much less efficient for P2 than for P1. One singular feature is the small difference (1 km) between the maximum injection heights of OC/P1 and OC/P2. It may be related to the emission rate increase between P1 and P2 which is the strongest for OC compared to CO or BC.

To the right of the longitudinal cross-sections of Figure 10 we plot the vertical distribution of both P1 and P2 above Madrid on 8 September at 00UT and superimpose the particle backscatter coefficient measured in Madrid on 7 September at 21 UT. In order not to rely on a single profile of the simulations, P1 and P2 profiles are averaged over a square of 9 pixels centered around the coordinates of Madrid. At time of arrival over the IP, HYSPLIT results for all compounds contribute to assign Plume 2 as the main source of smoke particles, representing more than 90 % of the column density. At arrival over the IP CO is present up to 13 km, BC up to 11 km and OC up to 12 km (Figure 10, right plots). The concentration of all three compounds is low in the first height interval between 0 and 2.5 km, except for CO/P1. For CO and BC the concentration levels of P1 are higher than for P2 which supports a former suggestion (see Section 4) that the airmasses arriving above the IP at 3 km on 8 September picked up smoke most likely from Plume 1, while those arriving at 6 and 11 km most likely from Plume 2. The peak of CO near 5-km height is very well reproduced by the model: it matches exactly the peak of the backscatter coefficient. For BC and OC the concentration peak (at 4 km) is 1 km lower than the peak of the backscatter coefficient (at 5 km). For both types of particles the gradual decrease of the concentration with increasing height above the peak reflects well the behavior of the backscatter coefficient. Given the poor model vertical resolution and the long distance of the horizontal transport (~10000 km), the particle transport is indeed very well simulated at its arrival in the IP. HYSPLIT simulates the presence of a layer of BC at ~11 km in the upper troposphere and a layer of OC at 10 – 12 km just below the tropopause, whereas the observation indicates that the smoke plume is not present above 10 km. Interestingly enough if the fact that a tiny layer of BC is simulated by HYSPLIT above the tropopause at 14 km. With respect to the literature, the concentrations simulated by HYSPLIT correspond to relatively small amount of what is usually measured at ground level. In a city like Barcelona, where BC is abundantly produced, the background BC concentration is usually higher than 1000 ng m$^{-3}$ (Pérez et al., 2010), i.e. much higher than the values simulated after long-range transport which peak at 10 ng m$^{-3}$.

Finally, we now come back to some of the hypotheses made in Section 5.2 and look for supporting arguments with the results of the dispersion modelling. With respect to the material burnt possibly containing low carbon content or the dominance of smoldering combustion, the dispersion modelling is of no help. The lower lidar ratio at 532 nm in the upper troposphere compared to the mid troposphere reflects less absorbing particles and possibly a lesser amount of BC with respect to OC. This tendency is actually confirmed by the vertical distributions of BC and OC at their arrival over the IP: BC and OC peak at 4 km and then gradually decrease up to 12 km, and the relative decrease of BC is stronger than for OC. About the increase of the depolarization ratio with height, the dispersion modelling is of little help, the result the most useful being that in the upper troposphere only P2 is present. Overall the observations and the dispersion modelling point out to the following: near the source the smoke particles slightly depolarize ($\delta_p = 0.05$ at ~6 km height, CALIOP, D-5) and at the arrival over the IP after a 5-day transport the particles have gained altitude and $\delta_p$ has increased ($\delta_p = 0.10$ at ~12 km, Évora, D-0). The particles arriving over the IP at ~6 km have unchanged

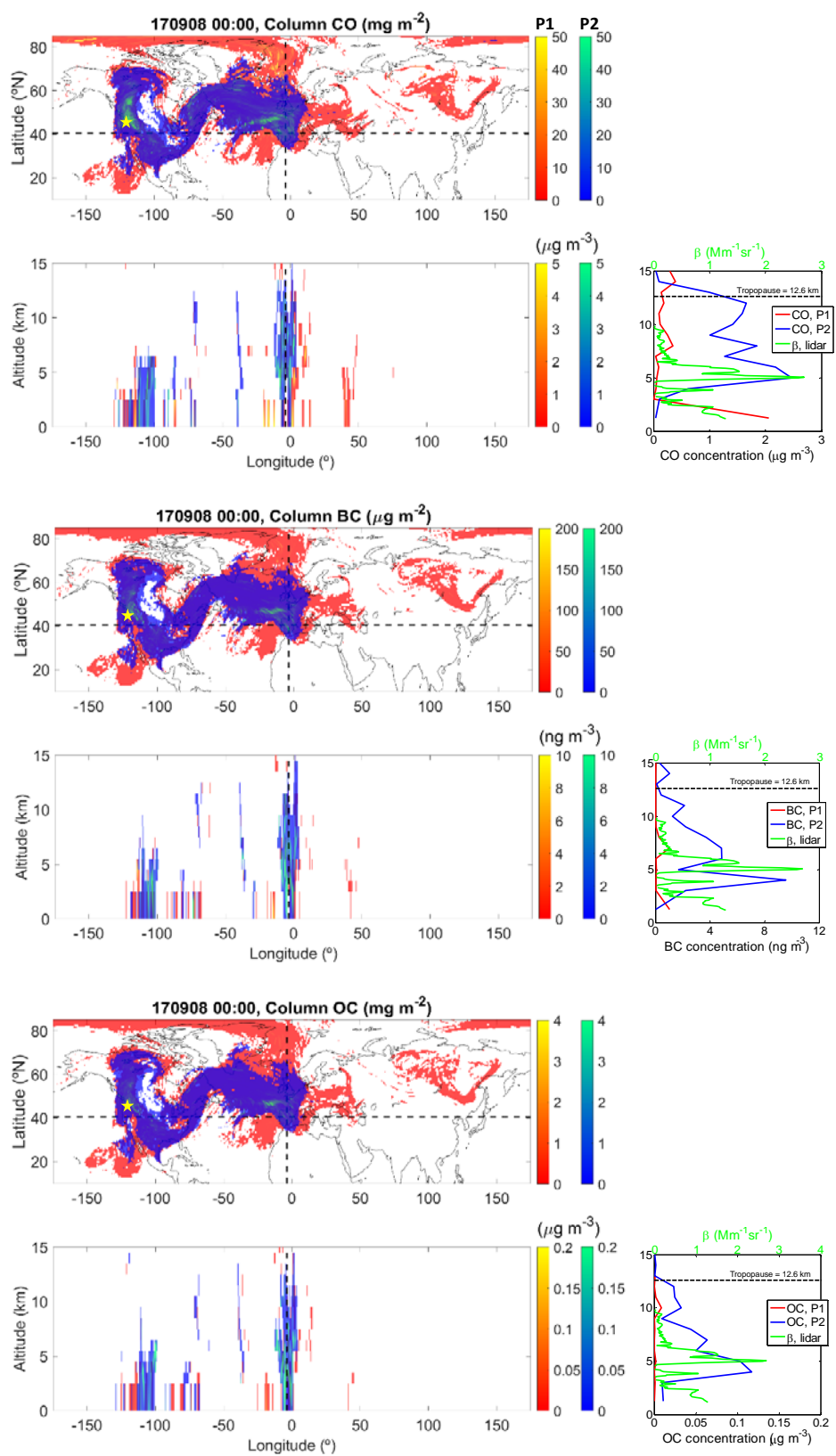

**Figure 10:** (top) Dispersion map of CO column density and longitudinal cross-section of CO concentration at the latitude of Madrid on 8 September at 00UT; (center) the same for BC; (bottom) the same of OC. Note the different scales. The emission and dispersion

of P1 (red-yellow color bar) and P2 (blue-green color bar) are separated. To the right of the longitudinal cross-sections we report the vertical profile of each chemical compound at the coordinates of Madrid for P1 and P2, as well as the backscatter coefficient at 532 nm retrieved in Madrid on 7 September at 21UT. The yellow star indicates the fire source region.

depolarization properties ($\delta_p = 0.05$) with respect to CALIOP, D-5. These findings enlighten the enhancement of the smoke depolarization properties with vertical transport. As smoke particles are relatively effective cloud condensation nuclei (Reid and Hobbs, 1998; Warner and Twomey, 1967), we finally hypothesize that smoke particles at non-dry altitude levels such as the upper troposphere (relative humidity ~ 20-30 %, see Section 5.2) may suffer freezing which may accentuate their asymmetric form and thus their depolarization properties.

# 7 Conclusions

This paper documents the time-space evolution of a smoke plume detected at its arrival over the Iberian Peninsula on 7 and 8 September, 2017. The smoke was emitted by strong and powerful wildfires in the Pacific northwestern region of North America, a region mostly composed of temperate coniferous forests. The column properties retrieved at two mid-altitude, background AERONET sites in northern and southern Spain reveal $AOD_{440}$ as high as 0.62, exceeding the background AOD by a factor larger than 6, $AE_{440-870}$ of 1.6-1.7, a large dominance of small particles ($FMF > 0.88$), low $AAOD_{440}$ (<0.008) and large $SSA_{440}$ (>0.98). The low absorption properties are attributed either (i) to the burning of low carbon content fuels such as forest floors, particularly abundant in temperate forests, (ii) the dominance of smoldering vs. flaming combustion, and/or (iii) a transformation (coating processes) of the smoke particles during transport. AAE ~ 1.3, together with large $SSA_{440}$, in northern Spain is representative of brown carbon, while AAE ~ 1.0, also associated with large $SSA_{440}$, in southern Spain is representative of brown carbon probably mixed with pure BC from the anthropogenic fossil fuel emissions of the nearby city of Granada, or from local fires approximately 150 km northwest of the site.

Satellite images of total column CO allows to identify two strong periods of emission that gave birth to two different plumes reaching the IP almost simultaneously: Plume 1 is emitted from 30 August until 1 September and Plume 2 from 2 to 5 September. The vertical distribution of the smoke plumes was monitored by ground-based lidars from both EARLINET and MPLNET networks, and from space by CALIOP. Over the IP a continuum of aerosols is observed up to the upper troposphere: aerosols are present in the whole troposphere. No particles are observed in the low stratosphere. Results are given for the mid (5 – 9 km) and upper (10 – 13 km) troposphere. The analysis of the ground-based lidars indicates a color ratio of 2.5 (3.0), $LR_{532}$ of 55 (34) sr, and $\delta_p$ of 0.05 (0.10) in the mid (upper) troposphere, which points out to smaller, less absorbing and more depolarizing particles in the upper troposphere than in the mid troposphere. Rewinding in time with CALIOP, one observes that the older the smoke plume, the larger the color ratio, i.e. that the particle size gets smaller during transport. As far as the particle depolarization ratio is concerned, no changes related to the transport are observed in the mid troposphere. The unusual values of $\delta_p$ in the upper troposphere (0.10) are further analyzed with dispersion modelling.

To analyze the horizontal and vertical transport of the smoke from its origin to the IP, particle dispersion modelling is performed with HYSPLIT parameterized with satellite-derived biomass burning emission estimates from GFAS/CAMS. We simulated CO, BC and OC, for separately P1 and P2, with a time resolution of 6 hours, at 15 altitude levels and using meteorology data from GDAS with a horizontal resolution of 0.5 x 0.5º. The smoke release height was not artificially fixed, but calculated internally by the model and assumed to be equal to the final buoyant plume rise height as computed using Briggs (1969), implying that the final rise is a function of the input fire radiative power and the meteorology. The results show that the dispersion of both plumes is quite different: P1 travels slowly and disperses over a large area of the northern hemisphere, while P2 is entrained by a strong subtropical jet and travels quickly towards the IP. The ascending rate of CO is nearly twice larger for P1 than for P2: CO/P1 reaches the height of 11 km in 36 hours, while CO/P2 needs 60 hours. There is undeniably a tradeoff between vertical and horizontal transport: slow horizontal transport is favorable to vertical motion whereas strong horizontal transport reduces it. At time of arrival over the IP, both BC and OC profiles over the IP are similar in shape to the lidar-derived backscatter coefficient profile: they both peak at 4 km and then gradually decrease up to 12 km, and the relative decrease of BC is stronger than for OC, which corroborates one of the former hypothesis, namely that particles in the upper troposphere are less absorbing than in the mid troposphere because of a smaller ratio of BC to OC. HYSPLIT results for all compounds contribute to assign P2 as the main source of smoke particles over the IP, representing more than 90 % of the column density. These findings, all together, show that $\delta_p$ increase from 0.05 to 0.10 occurs during the vertical transport from the mid to the upper troposphere, and stress the influence of the vertical transport on the smoke depolarization properties. As smoke particles are relatively effective cloud condensation nuclei, we finally hypothesize that smoke particles at non-dry altitude levels such as the upper troposphere may suffer freezing which may accentuate their asymmetric form and thus their depolarization properties.

**Acknowledgments**

This work was supported by the European Union through H2020 programme (ACTRIS-2, grant 654109, EUNADICS-AV, grant 723986; GRASP-ACE, grant 778349), and the European Fund for Regional Development (ref. POCI-01-0145-FEDER-007690, ALT20-03-0145-FEDER-000004, ALT20-03-0145-FEDER-000011 and 0753_CILIFO_5_E). Spanish groups acknowledge the Spanish Ministry of Economy and Competitivity (MINECO) (ref. CGL2013-45410-R, CGL2014-52877-R, CGL2014-55230-R, TEC2015-63832-P, CGL2015-73250-JIN, CGL2016-81092-R and CGL2017-85344-R), the Spanish Ministry of Sciences, Innovation and Universities (ref. CGL2017-90884-REDT), and the Unidad de Excelencia María de Maeztu (grant MDM-2016-0600) funded by the Agencia Estatal de Investigación, Spain. This work was also supported by the Juan de la Cierva-Formación program (grant FJCI-2015-23904), the MINECO Programa de Ayudas a la Promoción del Empleo Joven e Implantación de la Garantía Juvenil en i+D+i (grant PEJ-2014-A-52129), the Spanish Ministry of Education, Culture and Sport (MECD) (grant FPU14/03684), and the Portuguese Foundation for Science and Technology and national funding (ref. SFRH/BSAB/143164/2019). Co-funding was also provided by the Andalusia Regional Government (grant P12-RNM-2409), by the University of Granada through "Plan Propio. Programa 9 Convocatoria 2013", and by the Madrid Regional Government (projects TIGAS-CM, ref. Y2018/EMT-5177 and AIRTEC-CM, ref. P2018/EMT4329). The MPLNET project is funded by the NASA Radiation Sciences Program and Earth Observing System. The authors gratefully acknowledge the NOAA Air Resources Laboratory (ARL) for the provision of the HYSPLIT transport and dispersion model and/or READY website (http://www.ready.noaa.gov), and the CALIPSO mission scientists and associated NASA personnel for the production of the data used in this research.

List of Figure Captions

Figure 1: MODIS/Aqua corrected reflectance (true color) map centered over Spain on 8 September. Green bullets indicate lidar stations (EV: Évora, AR: El Arenosillo/Huelva, GR: Granada, MA: Madrid, BA: Barcelona) and red bullets indicate AERONET sites. Map created from https://firms.modaps.eosdis.nasa.gov/map/.

Figure 2: Flowchart of the methodology.

Figure 3: AOD440 (black), FMF (blue) and AE440-870 (red) in (top) Montsec, northeastern Spain, and (bottom) Cerro Poyos, south Spain. The gray areas in the bars on top of the figures indicate coincident lidar measurements.

Figure 4: Total column carbon monoxide (day/night) from AIRS/AQUA from 30 August until 8 September. The extra plot at the bottom to the right represents the MODIS combined (Aqua and Terra) value-added AOD at 550 nm on 8 September. The red star indicates the position of the active fires. On the plots of 3 and 4 September the descending, nighttime orbits of CALIPSO are reported. Maps created from https://worldview.earthdata.nasa.gov/.

Figure 5: (top) 10-day back-trajectories, 1-hour resolution, arriving in Madrid, in the center of Spain, on 8 September at 00UT at heights of 3 (red), 6 (green) and 11 (blue) km; (bottom) Same back-trajectories, different viewing angle and superposition of CALIOP curtains on 4 September at 05:10UT (D-4, day-4 before arrival) and on 3 September at 09:23UT (D-5) where the smoke plumes, clearly visible, match very well in space and time with the back-trajectories. Pink crosses indicate active fires in the period 30 August – 5 September. The red rectangle of corner coordinates (125W, 40N; 93W, 58N) is the area in which the fires were taken into account in the dispersion modelling analysis (see Section 6). The orange rectangle simply highlights the region containing most of the fires. Maps created with Google Earth.

Figure 6: AERONET daily mean spectral (top) AAOD, (center) SSA, and (bottom) asymmetry factor at Montsec and Cerro Poyos on 7 and 8 September.

Figure 7: Nighttime multi-wavelength lidar inversion in Évora on 7 September between 04 and 06UT. The first plot represents the quicklook of range-square corrected signal at 1064 nm in arbitrary units. $\beta$ is the particle backscatter coefficient, $\alpha$ the particle extinction coefficient, $\alpha-AE$ the extinction-related AE, $CR$ the color ratio, $LR$ the lidar ratio and $\delta_p$ the particle depolarization ratio. Mean values in the mid troposphere and stratosphere (as depicted by the gray rectangles) for $\alpha-AE$, $CR$, $LR$ and $\delta_p$ are reported in the plots. The horizontal dash lines at 13.6 km indicate the tropopause height calculated with 1º x 1º GDAS data.

Figure 8: CALIOP images and products on (left) 4 September at 05:10UT (D-4, Plume 1 released 5 days earlier) and (right) 3 September at 09:23UT (D-5, Plume 2, fresh < 1 day). (top) CALIOP quicklooks of the total attenuated backscatter signal at 532 nm; (center) CALIOP quicklooks of the retrieved backscatter coefficient at 532 nm restricted to the smoke plume (red squares); (bottom) CALIOP mean profiles of backscatter coefficient at 532 and 1064 nm, the color ratio and the particle depolarization ratio at 532 nm. The horizontal black dash lines indicate the tropopause height calculated with 1º x 1º GDAS data.

Figure 9: (top) Mid and upper tropospheric layer mean particle depolarization ratios at 532 nm at all Iberian lidar stations on the night of 7 to 8 September. Cyan and Purple bullets represent CALIOP measurements. The vertical bars indicate the vertical extension of the smoke layers of maximum intensity (base to top height). The horizontal bars indicate the standard deviation associated to $\delta_p$ in these layers. (bottom) Layer mean particle depolarization ratios at 532 nm vs. layer mean color ratio. The bullet color code is the same as in the top plot. We have reported four aerosols classes adapted from Groß et al. (2013). The vertical and horizontal bars indicate the standard deviation associated to $\delta_p$ and $CR$, respectively.

Figure 10: (top) Dispersion map of CO column density and longitudinal cross-section of CO concentration at the latitude of Madrid on 8 September at 00UT; (center) the same for BC; (bottom) the same of OC. Note the different scales. The emission and dispersion of P1 (red-yellow color bar) and P2 (blue-green color bar) are separated. To the right of the longitudinal cross-sections we report the vertical profile of each chemical compound at the coordinates of Madrid for P1 and P2, as well as the backscatter coefficient at 532 nm retrieved in Madrid on 7 September at 21UT. The yellow star indicates the fire source region.